\documentclass[11pt]{amsart}
\usepackage{epsfig,pgf,wrapfig,hyperref}
\usepackage{amssymb,latexsym,amsmath,color,cite,url}
\newtheorem{theorem}{Theorem}

\title{Parameter Estimation from ICC curves}
\author{Joceline Lega}
\address{Department of Mathematics, University of Arizona, 617 N. Santa Rita Avenue, Tucson, AZ 85721}
\keywords{Outbreak; Parameter Identification; Compartmental Models}

\begin{document}
\begin{abstract}
Incidence vs Cumulative Cases (ICC) curves are introduced and shown to provide a simple framework for parameter identification in the case of the most elementary epidemiological model, consisting of susceptible, infected, and removed compartments. This novel methodology is used to estimate the basic reproduction ratio of recent outbreaks, including the ongoing COVID-19 epidemic.
\end{abstract}

\maketitle

\section{Overview}
This article introduces the concept of ICC (Incidence  vs Cumulative Cases) curves, which are nonlinear transformations of the traditional `EPI curves' (plots of disease incidence versus time) commonly used by epidemiologists. The main message of the present work is that describing an outbreak in terms of its associated ICC curve is not only natural from a dynamical point of view, but also advantageous for parameter identification and forecasting purposes. Specifically, we provide a method for the practical identification of epidemiological parameters of the SIR (Susceptible - Infected - Removed) compartmental model, directly from its associated ICC curve. Because the result is robust to under-reporting, this approach gives a simple way to estimate the basic reproduction ratio of a disease from reported incidence data. The present analysis also provides a theoretical justification for the forecasting methodology put forward in \cite{Lega16}, which was shown to lead to useful estimates of the characteristics (expected duration, final number of cases, and peak intensity) of ongoing outbreaks. 

We develop the concept of ICC curves in the context of the classical SIR model \cite{Kermack27, Hethcote00}, which is the simplest compartmental model that captures the basic phenomenon of disease transmission: before recovering (or being removed through disease-induced death), infected individuals transmit the disease to susceptible individuals, who in turn become infected. The process repeats itself until the epidemic has followed its course and no additional infections occur. Since many outbreaks have a time scale that is much shorter than the scale at which the total population $N$ changes, a common simplification is to assume that $N$ is constant. It is then natural to define the time-dependent quantities $s = S/N,$ $i = I/N,$ and $r = R/N$ as the expected sizes of the susceptible ($S$), infected ($I$), and recovered ($R$) compartments relative to $N,$ so that $s + i + r = 1.$ The resulting SIR model is then given by 
\[
\frac{ds}{dt}= - \beta s i; \quad \frac{di}{dt}= \beta s i - \gamma i; \quad \frac{dr}{dt}= \gamma i,
\]
where $\beta$ is the contact rate of the disease and $\gamma$ is its recovery rate. Scaling time by $\tau = 1/\gamma$ turns the above equations into a one-parameter ($R_0$) model, where $R_0 = \beta / \gamma$ is the basic reproduction ratio \cite{Diekmann90,vandenDriessche02, Diekmann10} of the modeled epidemic. The disease spreads if $R_0 > 1$ and dies out when $R_0 < 1.$ It is also clear from the right-hand side of the second SIR equation that for $i \ne 0,$ the proportion of infected individuals grows as a function of time as long as $ s > \gamma / \beta = 1/R_0.$

\begin{figure}[h]
\includegraphics[width=\linewidth]{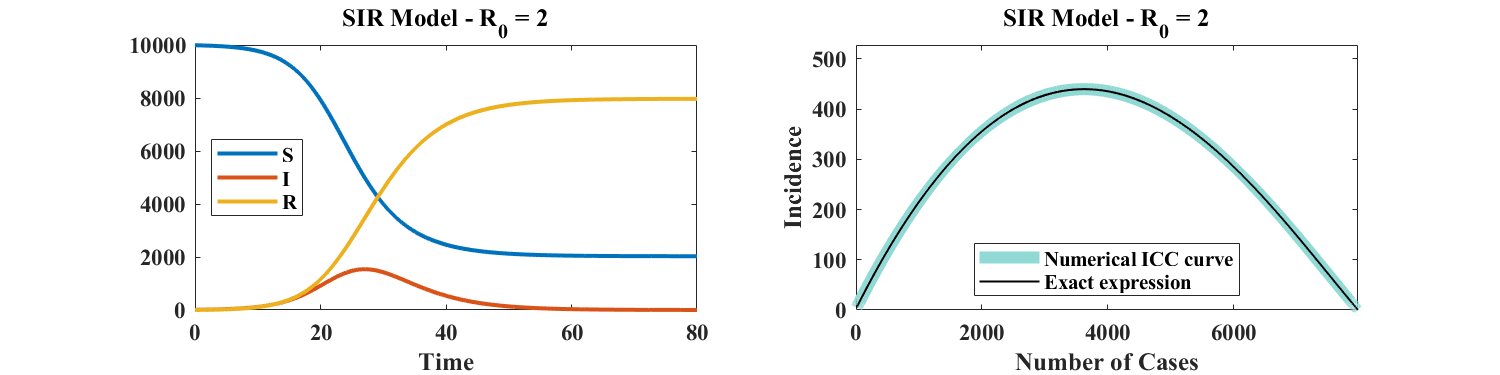}
\caption{\label{figure:SIR_simulations} (Color online). Simulation of the SIR model showing the time-dependence of the S, I, and R compartments (left), for $R_0 = 2$. The corresponding ICC curve is shown on the right. The exact expression given by Equation \eqref{eq:ICC} is plotted as a solid black line; the numerically estimated ICC curve is shown as a  thick cyan line.}
\end{figure}

In a compartmental model, an {\em outbreak} is a trajectory that starts in the vicinity of the disease-free equilibrium and ends at the endemic equilibrium, when the number of infected individuals approaches zero. We define an {\em ICC curve}, where ICC stands for Incidence vs. Cumulative Cases, as a representation of the dynamics associated with an outbreak, in the plane of coordinates $C$ (cumulative cases) and $\mathcal I$ (incidence). The following theorem, established in Section \ref{sec:ICC_curves}, gives an exact formulation of the ICC curves of the SIR model.
\begin{theorem}
For a total susceptible population of size $N$ and each initial condition $\displaystyle \kappa = \frac{S(0)}{N} = 1 - \frac{C(0)}{N}$, the ICC curve of the SIR model is given by
\begin{equation}
\label{eq:ICC}
G_{\kappa,N}( C) = \beta \left(C + \frac{N}{R_0} \ln\left(1 - \frac{C}{N}\right) - \frac{N}{R_0} \ln(\kappa) \right) \left(1 - \frac{C}{N} \right),
\end{equation}
where $C = I + R \in [0, C_\infty]$ and $C_\infty,$ the final number of cases, is the positive solution of the transcendental equation
\begin{equation}
\label{eq:C0}
C_\infty + \frac{N}{R_0} \ln\left(1 - \frac{C_\infty}{N}\right) - \frac{N}{R_0} \ln(\kappa) = 0.
\end{equation}
\end{theorem}
Details of the derivation of this theorem are given in Section \ref{ICC_derivation}. Its significance is illustrated in Figure \ref{figure:SIR_simulations}, which shows a numerical simulation of the SIR model and a plot of the corresponding ICC curve for $R_0 = 2$. Instead of considering the time dependence of $S$, $I$, and $R$ as shown in the left panel of Figure \ref{figure:SIR_simulations}, the course of the outbreak is captured by a single curve: a plot of incidence versus cumulative cases. As explained in Section \ref{ICC_derivation}, knowledge of the ICC curve is sufficient to reconstruct the temporal behavior of $S$, $I$, and $R$. In other words, the ICC curve contains the entire information associated with any outbreak described by the SIR model.


An important motivation for studying ICC curves is that they can easily be fitted to case data. Specifically, we show in Section \ref{ICC_fit} that, for fixed $N$, there exists a single set of parameters $(\beta, \gamma, \kappa)$ that minimizes the root mean square error between (noisy) simulated outbreak data and the ICC curve ${\mathcal I} = G_{\kappa,N}(C)$ given by Equation \eqref{eq:ICC}. Additionally, a numerical investigation indicates that such an estimation is robust to noise. As a consequence, the present approach provides a method to estimate the parameters of the SIR model, thereby making this model {\em practically identifiable}.

The rest of this manuscript is organized as follows. Section \ref{sec:ICC_curves} presents the derivation of Equations \eqref{eq:ICC} and \ref{eq:C0}, and discusses the robustness of ICC curves to systemic under- (or over-) reporting. Section \ref{sec:params} explains how model parameters may be retrieved from ICC curves and assesses the robustness of the proposed procedure to reporting noise. Section \ref{sec:convergence} briefly discusses parameter estimation as an outbreak unfolds. Section \ref{sec:examples} shows a few examples of application to real data, for outbreaks of gastroenteritis and COVID-19. Future extensions and applications of the methodology proposed in this article are reviewed in Section \ref{sec:conclusions}.

\section{ICC curves of the SIR Model}
\label{sec:ICC_curves}
\subsection{Proof of Theorem 1.1}
\label{ICC_derivation}
Epidemiologists use the \textit{epidemiological (EPI) curve}, a plot of incidence as a function of time, to visualize the course of an epidemic. In the SIR model, incidence is defined as
\[
{\mathcal I} = \beta S \frac{I}{N} = \frac{d C}{dt},
\]
where $C = R + I$ is the cumulative number of cases at time $t.$ Note that $C$ includes those who have recovered (and thus were a case at some point in time), as well as those who are currently infected. To derive Equation \eqref{eq:ICC}, write $ds/dr = - \beta s / \gamma = - R_0\, s,$ which can be integrated to give $s =\kappa \exp(- R_0\, r),$ where $\kappa \in (0, 1)$ is a constant that depends on initial conditions. Typically, $\kappa \simeq 1$ (see e.g. \cite{Britton05}) since for an outbreak in a naive population we normally have $r(0)=0,$ $i(0) << 1,$ and thus $\kappa = s(0) \simeq1.$ However here we do not make this assumption and keep the parameter $\kappa$ in the following discussion. With $c = C/N$ and $s = 1 - (i + r) = 1 - c,$ we obtain
\[
r = -\frac{1}{R_0} \ln(s/\kappa) = -\frac{1}{R_0} \ln((1 - c)/\kappa),
\]
and consequently
\[
\frac{d c}{dt} = \beta i s = \beta (c - r) (1 - c) = \beta (c + \frac{1}{R_0} \ln((1-c)/\kappa)) (1 - c).
\]
Moving back to dimensional variables gives the expression in Equation \eqref{eq:ICC}, i.e. ${\mathcal I} = d C / d t = G_{\kappa,N}(C)$.

An approximate expression of the SIR ICC curve, assuming $\kappa = 1$ and $C/N$ sufficiently small was given in \cite{Lega16}. Since $G_{\kappa,N} \ge 0$ on $[0, C_\infty]$, for a given initial condition $\kappa = 1 - C(0) / N$, the part of the ICC curve traced out as the outbreak unfolds will be such that $C(t) \ge C(0) > 0$. However, the ICC curve is well defined for $C$ in the range $[0, C_\infty]$, which is the definition we adopt here. Even though the time variable does not appear in the ICC curve, the value of $\kappa$ is not arbitrary: $C(0)$ is the value of $C$ when $R=0$, where $R$ represents the removed compartment.

ICC curves are particularly useful for forecasting the course of an outbreak because they contain information on quantities that are of interest to public health practitioners, namely the final number of cases ($C_\infty$) and peak incidence (${\mathcal I}_{max}$). The latter is simply the maximum of $G_{\kappa,N}$; the former may be found as follows. Since $C$ is the cumulative number of cases, it is a monotonic function of time and its derivative, $G_{\kappa,N}(C),$ must remain non-negative. For $C/N$ small,
\[
G_{\kappa,N}(C)  = G_{\kappa,N}(0) + C (\beta - \gamma + \gamma \ln(\kappa)) + {\mathcal O}(C^2)
\]
where $G_{\kappa,N}(0) = - N \gamma \ln(\kappa) > 0.$ By continuity, $G_{\kappa,N}(C)$ will remain non-negative on the interval $[0,\ C_\infty],$ where $C_\infty$ is the unique positive solution of Equation \eqref{eq:C0}. By writing Equation \eqref{eq:C0} as $N = C_\infty + \kappa N \exp(- R_0 C_\infty / N),$ where the right-hand-side is the sum of two non-negative monotonic functions of $C_\infty$, one increasing and the other decreasing, and since $0 < \kappa < 1,$ one can see that such a solution exists and is unique. Asymptotically, when the disease has followed its course and all infections have occurred, $I = 0$ and $C = R.$ Then, $S = \kappa N \exp(- R_0 R/N) = \kappa N \exp(- R_0 C / N)$ and the conservation law $N = S + I + R$ becomes $N = \kappa N \exp(- R_0 C / N) + C.$ In other words, $C_\infty$ defined in Equation \eqref{eq:C0} is simply the final (asymptotic) number of cases associated with the outbreak. This concludes the proof of Theorem 1.

To see that the entire dynamics of an outbreak in the SIR model is captured by the associated ICC curve, note that knowledge of $C(t)$ leads to knowledge of
\[
S(t) = N - C(t), \quad R(t) = - \frac{N}{R_0} \ln\left(\frac{1}{\kappa} \left(1 - \frac{C(t)}{N}\right)\right), \quad \text{and} \quad I(t) = C(t) - R(t).
\]
In other words, the time course of $S$, $I$, and $R$ may be obtained from the solution of the differential equation $d C / d t = G_{\kappa,N}(C)$ prescribed by the ICC curve, with appropriate initial conditions (which in turn define the value of $\kappa$). 

\subsection{Robustness to under- or over-reporting}
\label{sec:robustness}
The quantity $N$ appearing in Equation \eqref{eq:ICC} refers to the total population taken into account in the model, which is conserved by the dynamics. In case of over- or under-reporting, the observed incidence expectation ${\mathcal I}_o$ is a fraction of the actual incidence $\mathcal I$, so that ${\mathcal I}_o = \alpha {\mathcal I},$ where we assume that $\alpha$ is constant or varying slowly enough for its derivative to be negligible. The ICC curve estimated from case report data will then be a graph of ${\mathcal I}_o$ as a function of $\displaystyle C_o = \int_0^t {\mathcal I}_o(s) ds = \alpha\, C.$ We thus have
\begin{align*}
\frac{d {\mathcal I}_o}{d t} &= \alpha \frac{d \mathcal I}{d t} = \alpha\, G_{\kappa,N}( C) 
= \alpha \beta \left(C + \frac{N}{R_0} \ln\left(1 - \frac{C}{N}\right) - \frac{N}{R_0} \ln(\kappa) \right) \left(1 - \frac{C}{N} \right)\\
&  = \beta \left(C_o + \frac{\alpha  N}{R_0} \ln\left(1 - \frac{C}{N}\right) - \frac{\alpha N}{R_0} \ln(\kappa) \right) \left(1 - \frac{C}{N} \right)\\
& = \beta \left(C_o + \frac{N_o}{R_0} \ln\left(1 - \frac{C_o}{N_o}\right) - \frac{N_o}{R_0} \ln(\kappa) \right) \left(1 - \frac{C_o}{N_o} \right) = G_{\kappa,N_o}(C_o),
\end{align*}
where $N_o = \alpha N$. Therefore, the ICC curve in the presence of over-reporting ($\alpha > 1$) or under-reporting ($\alpha < 1$) has the same functional form, with the same parameter values, as the true ICC curve, provided $\mathcal I$, $C$, and $N$ are rescaled by the factor $\alpha$. For the same reason, $\kappa = S(0)/N$ is also independent of the value of $\alpha$.

While systematic under- or over-reportding does not affect the ICC curve provided the size of each compartment is appropriately rescaled,
the reproduction ratio depends on $N$ through the ratio $C_\infty/N$, as can be seen from Equation \eqref{eq:C0}, in which setting $\kappa = 1$ leads to
\begin{equation}
\label{eq:R0}
R_0 \frac{C_\infty}{N} + \ln\left(1 - \frac{C_\infty}{N}\right) = 0 \Longrightarrow
R_0 = - \frac{\ln\left(1 - C_\infty / N \right)}{C_\infty/N}.
\end{equation}
Finally, rescaling time amounts to rescaling $\beta$ and $\gamma$, and consequently the ICC curve. 

\section{Parameter identification}
\label{sec:params}
Structural identifiability (see e.g. \cite{Miao11} for a review) relates to the ability of uniquely identifying model parameters from knowledge of the model output. In this context, $N$ is assumed to be known since it represents the population of the system to which the model is applied. Given $C(t),$ one can calculate ${\mathcal I}(t) = dC / dt = G_{\kappa,N}(C(t)).$ From Equation \eqref{eq:ICC}, it is clear that knowledge of $C(t)$ and ${\mathcal I}(t)$ uniquely determines $\beta,$ $\gamma,$ and $\kappa.$ Indeed, if two sets of parameters, $(\beta^{(1)}, \gamma^{(1)}, \kappa^{(1)})$ and $(\beta^{(2)}, \gamma^{(2)}, \kappa^{(2)})$, led to the same output $C(t)$ (and thus to the same values of its derivative ${\mathcal I}(t)$), we would have 
\[
0 = \left(\beta^{(1)} - \beta^{(2)}\right) \frac{C}{N} + \left(\gamma^{(1)} - \gamma^{(2)}\right) \ln\left(1 - \frac{C}{N}\right) - \left(\gamma^{(1)} \ln(\kappa^{(1)}) - \gamma^{(2)} \ln(\kappa^{(2)})\right),
\]
for all values of $\frac{C}{N} \in \left[0, \frac{C_\infty}{N}\right]$, which leads to $\beta^{(1)} = \beta^{(2)}$, $R_0^{(1)} = R_0^{(2)}$, and $\kappa^{(1)} = \kappa^{(2)}.$ Therefore, all of the parameters of the SIR model, including initial conditions, are structurally identifiable from knowledge of $C(t),$ the cumulative number of cases, and $N$, the size of the susceptible population. Similar results were obtained in \cite{Evans05, Eisenberg13, Tuncer18}. If $N$ is unknown, setting 
\begin{align*}
0 = \beta^{(1)} \frac{C}{N^{(1)}} & - \beta^{(2)} \frac{C}{N^{(2)}} + \gamma^{(1)} \ln\left(1 - \frac{C}{N^{(1)}}\right) - \gamma^{(2)} \ln\left(1 - \frac{C}{N^{(2)}}\right) \\
& - \left(\gamma^{(1)} \ln(\kappa^{(1)}) - \gamma^{(2)} \ln(\kappa^{(2)})\right),
\end{align*}
for all values of $\frac{C}{N} \in \left[0, \frac{C_\infty}{N}\right]$, additionally leads to $N^{(1)} = N^{(2)}$, making the size of the population $N$ a structurally identifiable parameter as well.

Whereas structural identifiability depends on the nature of the model itself, practical identifiability relates to parameter identifiability in the presence of measurement noise, as well as (hopefully small) deviations between model and reality (see e.g. \cite{Miao11}). This question may be addressed numerically by simulating an exact solution of the model, adding noise to its output, and estimating the model parameters that best fit the perturbed output. Two recent articles \cite{Tuncer18,Roosa19} discuss the practical identifiability of a variety of compartmental models and reach different conclusions regarding parameter identification of the SEIR model from cumulative data. In \cite{Tuncer18}, the cumulative data is multiplied by $1 + \epsilon,$ where $\epsilon$ is normally distributed with zero mean, whereas in \cite{Roosa19}, Poisson-distributed noise of mean equal to the current incidence value \cite{Chowell17} is used as a substitute for the incidence data. Whether or not parameters are identifiable in practice therefore depends on the type of observational noise that affects surveillance reports and thus the EPI curve.

Adding independently-distributed noise to the incidence rather than the cumulative data is more realistic since errors on $C$ are considered to accumulate and not be independent \cite{King15}. On the one hand, assuming Poisson-type of uncertainty is natural for data that represent counts, but the effect of replacing incidence data by a Poisson random variable of same mean will lead to significant relative deviations only if $N$ is small (a few hundreds). For larger values of $N$ and thus of ${\mathcal I},$ the relative change $\delta{\mathcal I} = |{\mathcal I} - \hbox{Poisson}({\mathcal I})|/{\mathcal I}$ scales like $1/\sqrt{\mathcal I}.$ Indeed,
\begin{align*}
\langle \delta {\mathcal I} \rangle &= \sum_{k = 0}^{\mathcal I} \frac{{\mathcal I}^k e^{-{\mathcal I}}}{k !}
\frac{{\mathcal I} - k}{\mathcal I} + \sum_{k={\mathcal I} + 1}^\infty \frac{{\mathcal I}^k e^{- {\mathcal I}}}{k !}
\frac{k - {\mathcal I}}{\mathcal I}\\
&= \sum_{k = 0}^{\mathcal I} \frac{{\mathcal I}^k e^{- \mathcal I}}{k !}
- \sum_{k = 0}^{{\mathcal I} - 1}  \frac{{\mathcal I}^k e^{- \mathcal I}}{k !} + \sum_{k = {\mathcal I}}^\infty
\frac{{\mathcal I}^k e^{- \mathcal I}}{k !} - \sum_{k = {\mathcal I} + 1}^\infty  \frac{{\mathcal I}^k e^{- \mathcal I}}{k !}
= 2 \frac{{\mathcal I}^{\mathcal I} e^{-\mathcal I}}{{\mathcal I} !}.
\end{align*}
With Stirling's formula, we obtain
\[
\langle \delta {\mathcal I} \rangle  \sim 2 {\mathcal I}^{\mathcal I} e^{- \mathcal I} \left(\sqrt{2 \pi {\mathcal I}} {\mathcal I}^{\mathcal I} e^{- \mathcal I}\right)^{-1} = \sqrt{\frac{2}{\pi {\mathcal I}}} \qquad \hbox{ as } {\mathcal I} \to \infty.
\]
If it is believed that the diminishing relative influence of the Poisson-distributed noise for large values of $N$ is not realistic for a specific epidemic, then multiplying $\mathcal I$ by $(1 + \epsilon)$ where $\epsilon$ is normally distributed with mean zero, circumvents this issue. We consider both types of noise below.

\begin{figure}[h]
\includegraphics[width=\linewidth]{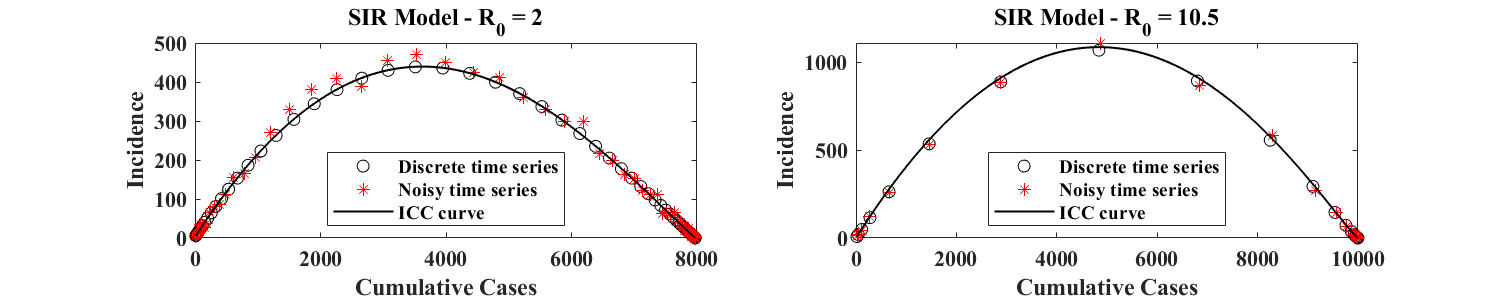}
\caption{\label{figure:ICC_noise} (Color online). Empirical ICC curves for two hypothetical outbreaks described by the SIR model with Poisson-distributed reporting noise. Incidence information is assumed to be available every $\delta_t$, with $\delta_t = 1$ for $R_0 = 2$ and $\delta_t =2$ for $R_0 = 10.5$. In each panel, stars correspond to the simulated outbreak data plotted in the ($C$, $\mathcal I$) plane. Data points without added noise (circles) and the exact ICC curve (solid line) are also shown.}
\end{figure}

To test practical identifiability, we simulate the SIR model and generate a discrete time series for the cumulative number of cases of the simulated outbreak. This time series is used as a reference and represents the unperturbed (or true) data. We then add noise to the associated discrete incidence data (defined as the difference of cumulative values between consecutive time points), following the procedure described in \cite{Chowell17}, with the additional realistic constraint that the final number of cases is the same for all noisy realizations of the same outbreak. Two examples of simulated outbreaks generated in this fashion are shown in Figure \ref{figure:ICC_noise} for different values of $R_0$ (associated with different orders of magnitude of the incidence $\mathcal I$) and different discretization time steps $\delta_t$. For $R_0 = 2$, we use $\delta_t = 1$. For $R_0 = 10.5$, we use $\delta_t = 2.$ As shown in Figure \ref{figure:ICC_noise}, this latter choice leads to a discretized ICC curve with very few points in regions of large incidence, and simulates a situation where incidence data is of poorer frequency. Each time series, consisting of noisy incidence data, is used to calculate an associated time series for the corresponding cumulative data. The resulting set of points is plotted in the ($C$, $\mathcal I$) plane, and defines an empirical ICC curve. In both panels of Figure \ref{figure:ICC_noise}, $C$ and $\mathcal I$ are estimated from discrete incidence data as described in Equation \eqref{eq:disc} below.

\subsection{Fitting ICC curves to data}
\label{ICC_fit}
Given a simulated time series for the number of cases of an outbreak in a population of size $N$, we find the parameters of the ICC curve that best fits the data by least square minimization. This approach is different from parameter identification methods that have been proposed in the literature until now. Indeed, traditional methods aim to find the best combination of model parameters consistent with observed time series; they therefore require numerical integration of the compartmental model whose parameters need to be identified, in order to estimate and then minimize an associated cost function (see for instance \cite{Chowell17, Tuncer18, Roosa19}). Instead, the approach we put forward here consists in finding model parameters by directly fitting the relevant ICC curve to the data. We claim that this is a more robust approach because the functional that needs to be minimized has a unique extremizer in parameter space, which can be calculated exactly from the data points.

Given a time series of discrete incidences $\{{\mathcal I}_k, k = 0, \cdots, M\},$ or of discrete cumulative cases $\{{\mathcal C}_k = \sum_{j=0}^k {\mathcal I}_j, k = 0, \cdots, M\}$ recorded every $\delta_t$ units of time, we introduce a cost function that measures how much the data points depart from the ICC curve of the SIR model parametrized by $\beta,$ $\gamma,$ and $p = \ln(\kappa).$ This function, ${\mathcal E}_e$, is defined by
\begin{equation}
\label{eq:cost}
{\mathcal E}_e(\beta,\gamma,p) = \sum_{i=1}^M \left(\frac{1}{N} {\mathcal I}\big(t_{i - \frac{1}{2}}\big) - G\big(\beta,\gamma,p,C\big(t_{i - \frac{1}{2}}\big)\big)\right)^2,
\end{equation}
where
\[
G(\beta,\gamma,p,C) = \left(\beta \frac{C}{N} + \gamma \ln\left(1 - \frac{C}{N}\right) - \gamma \, p \right) \left(1 - \frac{C}{N} \right),
\]
and
\begin{equation}
\label{eq:disc}
{\mathcal I}\big(t_{i - \frac{1}{2}}\big) = \frac{{\mathcal C}_i- {\mathcal C}_{i-1}}{\delta_t} = I_i, \quad C\big(t_{i - \frac{1}{2}}\big) = \frac{1}{2} \big({\mathcal C}_i + {\mathcal C}_{i-1}\big) = C_i.
\end{equation}
A Maple \cite{Maple} calculation shows that the cost function ${\mathcal E}_e$ has a unique critical point, whose expression is given in Appendix \ref{AppA}. For outbreaks that start in a naive population, the formulas are hugely simplified by setting $p = 0$. The resulting cost functional then has a unique minimum, whose value may easily be calculated. With the following notation,
\[
{\mathcal J}_i = \frac{I_i}{N}, \quad U_i = \frac{C_i}{N} \left(1- \frac{C_i}{N}\right), \quad V_i = -  \left(1-\frac{C_i}{N}\right) \ln \left(1-\frac{C_i}{N}\right),
\]
we have (after setting $p = 0$),
\[
\frac{\partial {\mathcal E}_e}{\partial \beta} = - 2 \sum_{i=1}^M U_i \left({\mathcal J}_i - \beta U_i + \gamma V_i\right); \quad \frac{\partial {\mathcal E}_e}{\partial \gamma} = 2 \sum_{i=1}^M V_i \left({\mathcal J}_i - \beta U_i + \gamma V_i \right).
\]
Setting the right-hand sides to zero and solving the resulting linear system for $\beta$ and $\gamma$ leads to the following expression for the unique global minimizer of ${\mathcal E}_e$ when $p = 0$:
\begin{align*}
\beta_m & = \displaystyle \frac{1}{h}\left(\left(\sum_{i=1}^M V_i^2\right) \left(\sum_{i=1}^M U_i {\mathcal I}_i\right) - \left(\sum_{i=1}^M V_i {\mathcal I}_i \right) \left(\sum_{i=1}^M U_i V_i\right) \right),\\
\gamma_m & = \frac{1}{h} \left(\left(\sum_{i=1}^M U_i V_i \right) \left(\sum_{i=1}^M U_i {\mathcal I}_i\right) - \left(\sum_{i=1}^M U_i^2\right) \left(\sum_{i=1}^M V_i {\mathcal I}_i \right) \right),
\end{align*}
where
\[
h = \left(\sum_{i=1}^M U_i^2\right) \left(\sum_{i=1}^M V_i^2\right) - \left(\sum_{i=1}^M U_i V_i\right)^2 = \frac{1}{2} \, \sum_{i,j = 1}^M \big(U_i V_j - V_i U_j \big)^2 > 0.
\]
In what follows, we use the expressions given in Appendix \ref{AppA}, which can easily be coded up. We checked that for time series generated from an SIR model with $\kappa \simeq 1$, the simplified expressions above provide good estimates of the actual values of $\beta$ and $\gamma$ that were used to generate the reference outbreak. Importantly, we now have a means of associating a single set of values for the parameters $\beta$, $\gamma$, and $p = \ln(\kappa)$ to each outbreak time series.

\subsection{Parameter identification in the presence of noise}
\label{sec:noise}
We now use the expressions found in Section \ref{ICC_fit} to assess the robustness of the proposed parameter identification method to noise. As discussed above, the SIR model will be declared practically identifiable provided good estimates of the parameters may be obtained from noisy data. We therefore simulate 100,000 noisy time series from a `reference' numerical integration of the SIR model, use the formulas of Appendix \ref{AppA} to calculate $p_c$, $\beta_c$, and $\gamma_c$ for each time series, and plot histograms of the estimated values of $\beta$ and $\gamma$. We consider the two types of noise discussed above, both affecting the reported discrete incidence $I_i = {\mathcal I}(t_{i - \frac{1}{2}}).$ In Case A, we assume that at each time point $t_{i - \frac{1}{2}},$ the reported value of $\mathcal I$ has a Poisson distribution of mean ${\mathcal I}_{ref}(t_{i - \frac{1}{2}}).$ In Case B, it is assumed that the reported value of $I_i$ at each time point is of the form $(1 + \epsilon s_i)\, {\mathcal I}_{ref}(t_{i - \frac{1}{2}}),$ where ${\mathcal I}_{ref}$ is the discrete incidence of the reference outbreak, $\epsilon$ is the noise amplitude, and $s_i$ is standard normal.

\subsubsection{Case A: Poisson-distributed incidence}
\begin{figure}[h]
\includegraphics[width=\linewidth]{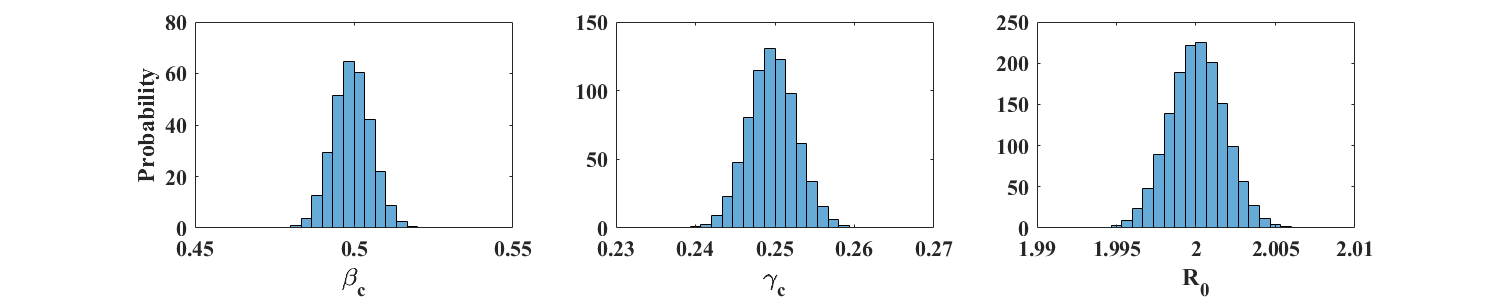}
\caption{\label{figure:param_dist_2} (Color online). Numerically estimated distributions of the parameter values $\beta_c$, $\gamma_c$, and $R_0 = \beta_c / \gamma_c$ for 100,000 outbreaks corresponding to noisy (Poisson-distributed) realizations of a reference outbreak with parameters $\beta = 0.5$ and $\gamma = 0.25$  ($R_0 = 2$). Corresponding means and standard deviations are displayed in Table \ref{table:T-dist}.}
\end{figure}

\begin{figure}[h]
\includegraphics[width=\linewidth]{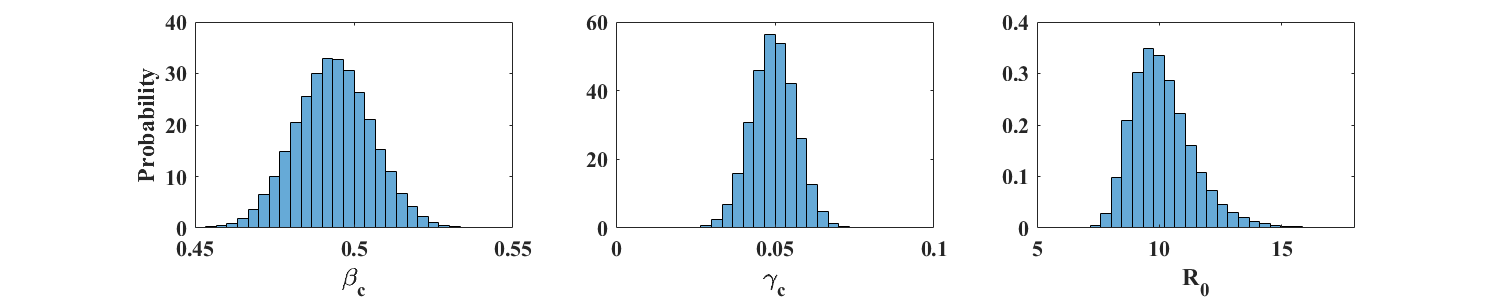}
\caption{\label{figure:param_dist_10.5} (Color online). Same as Figure \ref{figure:param_dist_2}, but with parameters $\beta = 0.5$ and $R_0 = 10.5$ ($\gamma \simeq 0.0476$). Corresponding means and standard deviations are displayed in Table \ref{table:T-dist}.}
\end{figure}

Figure \ref{figure:param_dist_2} shows the distributions (in the form of properly scaled histograms) of $\beta_c,$ $\gamma_c,$ and $R_{0} = \frac{\beta_c}{\gamma_c}$ obtained from 100,000 noisy realization from a reference simulation of an SIR model with parameters $\beta = 0.5$ and $\gamma = 0.25$ ($R_0 = 2$). Figure \ref{figure:param_dist_10.5} shows similar plots for $\beta = 0.5$ and $\gamma \simeq 0.0476$, so that $R_0 = 10.5$. In each case, the size of the total population is $N = 10000$ individuals. The corresponding sample means and standard deviations are displayed in Table \ref{table:T-dist}. These results indicate that the respective means of $\beta_c$ and $\gamma_c$ provide very good estimates of the parameters $\beta$ and $\gamma$. Accuracy is not as good for the larger value of $R_0$, whose distribution shows a fairly long tail. This is to be expected because the actual value of $\gamma$ in this case is close to zero. Moreover, the number of data points away from the end points of the outbreak is much smaller for $R_0 = 10.5$ than for $R_0 = 2$  (see Figure \ref{figure:ICC_noise}), leading to a potential loss of information.

\begin{table}[h]
{\begin{tabular}{ |c||c|c||c|c|}
 \hline
 & \multicolumn{2}{|c||}{$R_0 = 2$} & \multicolumn{2}{|c|}{$R_0 = 10.5$}\\
 \cline{2-3} \cline{4-5}
 & $\mu$ & $\sigma$ & $\mu$ & $\sigma$ \\
\hline
$\beta_c$  & 0.4994  & 0.0060  & 0.4936  & 0.0120 \\
\hline
$\gamma_c$ & 0.2497  & 0.0030 & 0.0495  & 0.0070\\
\hline
$R_0$ & 2.0001  & 0.0017  & 10.15  & 1.29 \\
 \hline
\end{tabular}}
\medskip
\caption{Empirical mean ($\mu$) and standard deviation ($\sigma$) of estimated parameter values in the presence of Poisson-distributed noise, for two values of $R_0$.
\label{table:T-dist}
}
\end{table}

\subsubsection{Case B: Normally-distributed noise}
At each time point, we now multiply the reference incidence by $1 + \epsilon s$, where $s$ is standard normal and the amplitude $\epsilon$ takes on one of 3 values: 0.05, 0.15, or 0.25. This type of noise is less realistic for incidence reports, whose variability is expected to be well described by a Poisson distribution, as in Case A above. We however perform the present tests to illustrate the robustness of the parameter estimation method introduced in this article.

\begin{table}[h]
{\begin{tabular}{ |c||c|c||c|c||c|c|}
 \hline
   & \multicolumn{6}{|c|}{$R_0 = 2$}\\
 \cline{2-7}
  & \multicolumn{2}{|c||}{$\epsilon=0.05$} &  \multicolumn{2}{|c||}{$\epsilon=0.15$} & \multicolumn{2}{|c|}{$\epsilon=0.25$}  \\
\cline{2-3} \cline{4-5}  \cline{6-7}
 & $\mu$ & $\sigma$ & $\mu$ & $\sigma$ & $\mu$ & $\sigma$ \\
\hline
$\beta_c$  & 0.4994  & 0.0048  & 0.5001  & 0.0152 & 0.5013 & 0.0267 \\
\hline
$\gamma_c$ & 0.2497  & 0.0024 & 0.2500  & 0.0076 & 0.2506 & 0.0135\\
\hline
$R_0$ & 2.0001  & 0.0009  & 2.0002  & 0.0028 & 2.0002 & 0.0050\\
 \hline \hline
  & \multicolumn{6}{|c|}{$R_0 = 10.5$}\\
 \cline{2-7}
& \multicolumn{2}{|c||}{$\epsilon=0.05$} &  \multicolumn{2}{|c||}{$\epsilon=0.15$} &  \multicolumn{2}{|c|}{$\epsilon=0.25$} \\
\cline{2-3} \cline{4-5}  \cline{6-7}
& $\mu$ & $\sigma$ & $\mu$ & $\sigma$ & $\mu$ & $\sigma$ \\
\hline
$\beta_c$  & 0.4939  & 0.0201  & 0.5053  & 0.0577 & 0.5396 & 0.0938 \\
\hline
$\gamma_c$ & 0.04961  & 0.0104 & 0.0550  & 0.0293 & 0.0717 & 0.0465\\
\hline
$R_0$ & 10.3682  & 2.14  & 27.0950  & 769.77 & 29.9396 & 863.86 \\
 \hline
\end{tabular}}
\medskip
\caption{Empirical mean ($\mu$) and standard deviation ($\sigma$) of estimated parameter values in the presence of normally-distributed noise of mean zero and amplitude $\epsilon$, for two values of $R_0$.
\label{table:N-dist}
}
\end{table}

\begin{figure}[h]
\includegraphics[width=\linewidth]{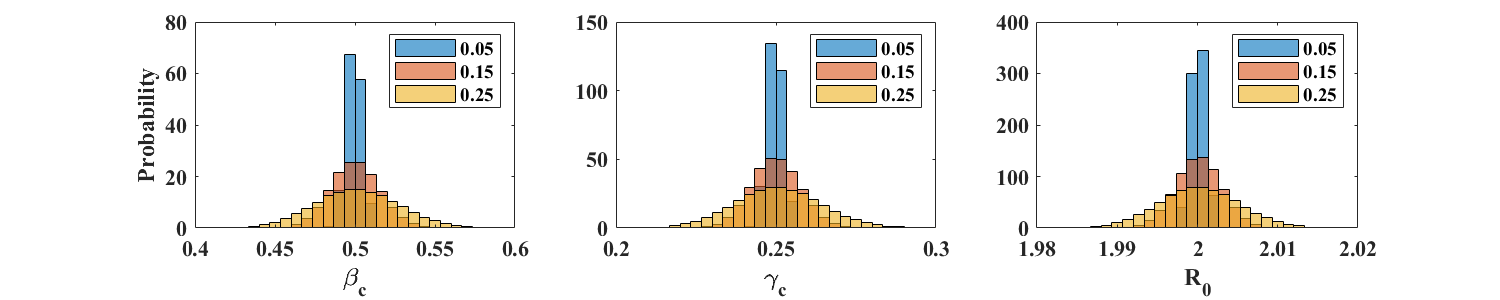} \\
\includegraphics[width=\linewidth]{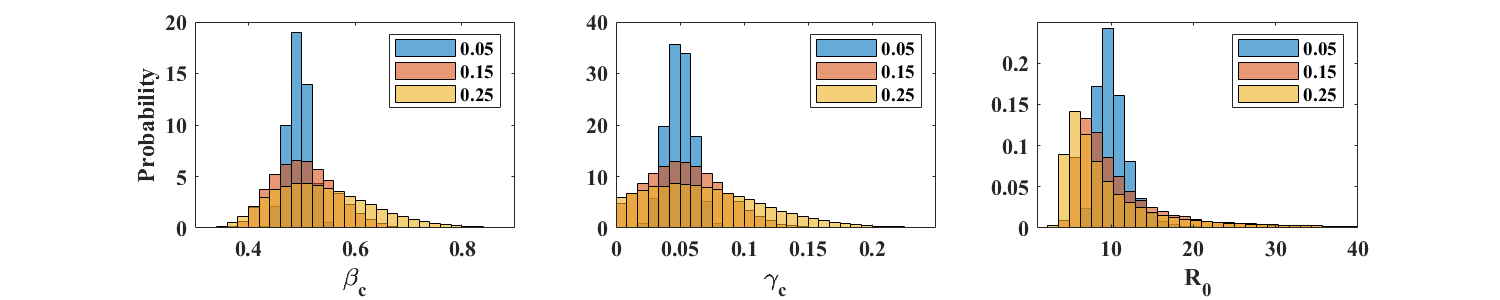}
\caption{\label{figure:param_dist_N} (Color online). Similar to Figures \ref{figure:param_dist_2} (top row) and \ref{figure:param_dist_10.5} (bottom row) but for normally-distributed noise of variable amplitude, set at 0.05, 0.15, or 0.25. Corresponding means and standard deviations are displayed in Table \ref{table:N-dist}.}
\end{figure}

For $R_0 = 2$, the parameter estimation procedure is robust at all values of $\epsilon$ considered. Means and standard deviations of $\beta_c$, $\gamma_c$, and $R_0 = \beta_c / \gamma_c$ are displayed in Table \ref{table:N-dist}. The corresponding distributions are shown in the top row of Figure \ref{figure:param_dist_N}. As expected, estimates of $\beta$ and $\gamma$ are not as accurate for simulations with $R_0 = 10.5$, but return mean values for $\beta$ and $\gamma$ that are close to the exact values. The distributions of $\beta_c / \gamma_c$ however shows very long tails  (see bottom row of Figure \ref{figure:param_dist_N}), due to values of $\gamma_c$ close to zero. In such a case, estimating $R_0$ as the ratio $<\beta_c> / <\gamma_c>$ however gives acceptable values, equal to 9.96, 9.19, and 7.53, for $\epsilon$ equal to 0.05, 0.15, and 0.25 respectively.

In summary, the above numerical simulations indicate that the ICC-based parameter estimation method presented here provides good results in the presence of reporting noise. Not surprisingly, the quality of the estimates depends on the rate at which incidence is sampled and, in the case of normally distributed noise, on the amplitude of the noise.

\section{Parameter identification as the outbreak unfolds}
\label{sec:convergence}
\begin{figure}[h]
\includegraphics[width=0.48 \linewidth]{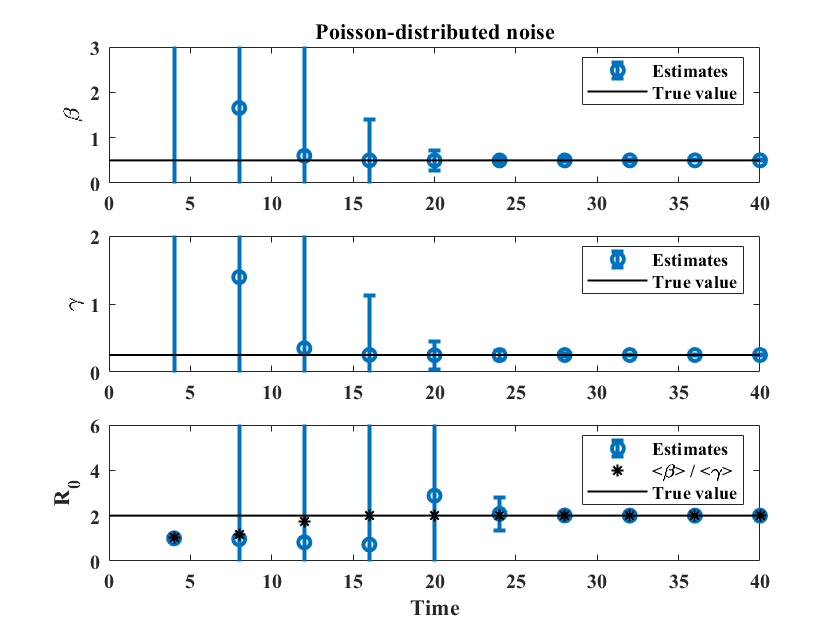}
\includegraphics[width=0.48 \linewidth]{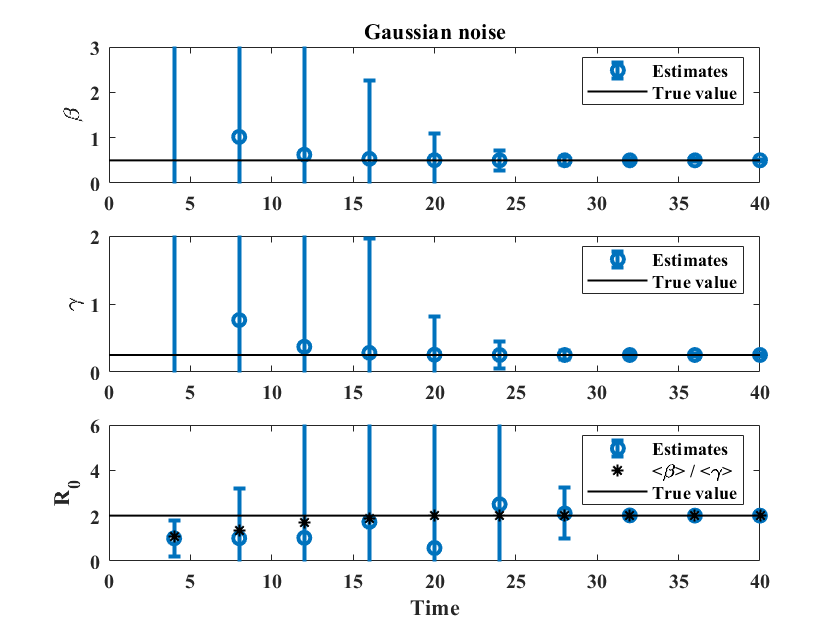}
\caption{\label{figure:convergence} (Color online). Estimates of $\beta$, $\gamma$ and $R_0$ as the outbreak unfolds for Poisson-distributed noise (left) and normally-distributed noise of size 0.15 (right), in a situation where $R_0 = 2$. Estimates of $R_0$ show the evolution of $<\beta / \gamma>$ (circles) and $<\beta> / <\gamma>$ (stars), where $< \cdot >$ indicates sample mean. Error bars correspond to one standard deviation on each side of the mean.}
\end{figure}

In principle, the formulas for $\beta_c$ and $\gamma_c$ given in Appendix \ref{AppA} may be used with any number of points $M$, as long as the data points represent the entire ICC curve. Is is natural to ask how the method performs when applied to a developing outbreak. To test this, we simulate 50,000 possible outbreaks using the same reference simulations as before, and with the two different types of noise introduced above to represent reporting error. We then follow the evolution of the distributions of the estimated parameters as time increases. For $R_0 = 2$, the outbreak has run its course after 60 units of time, and we include up to $M_{max} = 40$ data points, with consecutive values separated by 1 unit of time (which could correspond to a day or a week depending on the disease). For $R_0 = 10.5$, we pick $M_{max} = 80$, with data points separated by 2 units of time. Figure \ref{figure:convergence} shows estimated values of $\beta$, $\gamma$, and $R_0 = 2$ for the two types of reporting noise. In both cases, estimates of $\beta$ and $\gamma$ are close to their actual values near $t = 20$, which is before the peak of the outbreak (see left panel of Figure \ref{figure:SIR_simulations}). The value of $R_0$ converges more slowly (circles in the bottom row of Figure \ref{figure:convergence}) but the ratio of the estimated values of $\beta$ and $\gamma$ (stars) is close to the actual value of $R_0$ near $t = 12$, fairly early in the outbreak.

Similar results for $R_0 = 10.5$, together with plots showing the evolution of the distributions of $\beta_c$, $\gamma_c$, and $R_0$, are provided as supplementary material. Parameter estimates for $\beta$ and $\gamma$ are close to their true values near $t = 22$, which is near the peak of the outbreak, with the ratio $<\beta> / <\gamma>$ giving a reasonable estimate of $R_0$ starting at $t = 18$, even in the presence of large uncertainty.

\section{Application to outbreak data}
\label{sec:examples}
So far, we have assumed the size of the susceptible population, $N$, to be known. In practice, the value of $N$ that should be used to fit the ICC curve to outbreak data is often unknown since it may reflect under-reporting (as discussed in Section \ref{sec:robustness}) or spatial effects (existence of localized clusters to which the spread is restricted) associated with the outbreak. Moreover, it could be argued that the SIR model is too simplistic to represent real-life outbreaks. The examples of \cite{Lega16} and those discussed below however show there is merit to the present approach.

In what follows, we fit ICC curves to surveillance reports by picking a range of values of $N$ for which the root mean square error (RMSE) between the ICC curve and the data is within 2\% of the minimum RMSE value. The data points used in the evaluation of the RMSE are found from the reported cumulative data after smoothing and interpolation. This latter procedure estimates missing incidence values and removes negative incidence reports, if any. The resulting time series is expected to be a better approximation of the `true' (i.e. as close as possible to the output of a compartmental model) evolution of the cumulative data, and is used for the parameter estimation procedure discussed in this article. Our first example is a gastroenteritis outbreak in a nursing home in Mallorca, Spain, and is therefore spatially localized. The other examples are estimations of the basic reproduction ratio for COVID-19 outbreaks in Hubei Province (China), the Republic of Korea, and France, based on data available until 4/15/2020.

\subsection{Gastroenteritis outbreak in Mallorca, Spain}
\begin{figure}[h]
\includegraphics[width=\linewidth]{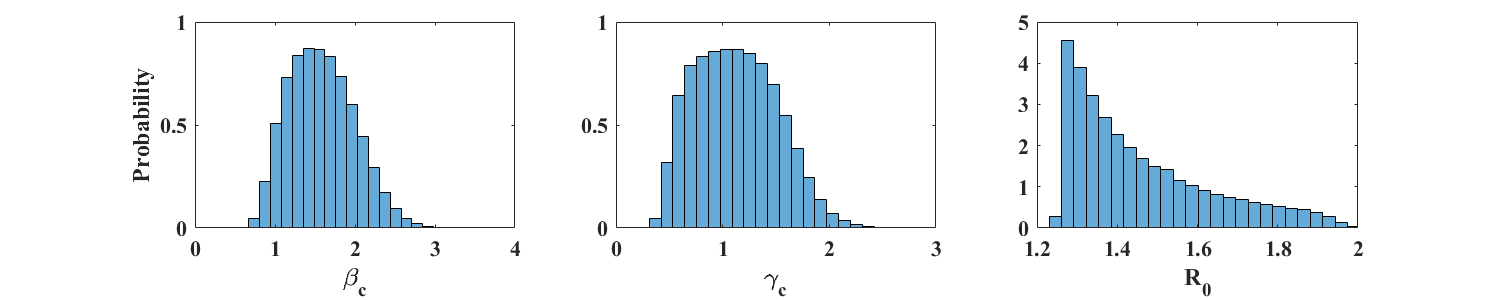} \\
\includegraphics[width=\linewidth]{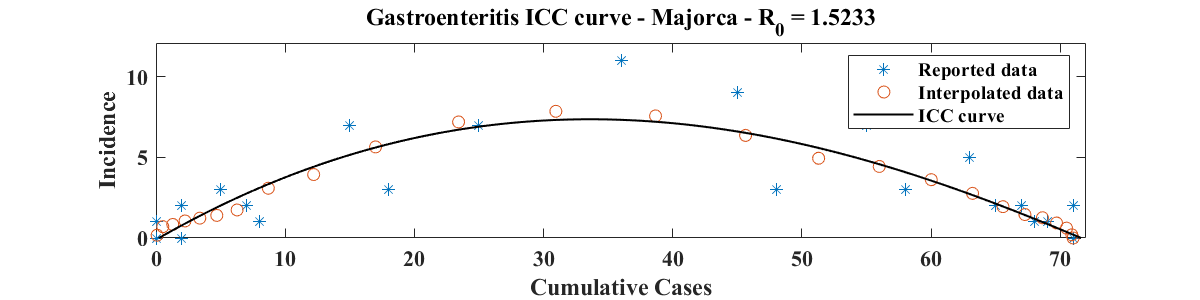}
\caption{\label{figure:Gastro} (Color online). Parameter estimates for a gastroenteritis outbreak in a nursing home in Mallorca, Spain \cite{Fernandez08}. The top panel shows histograms (scaled to represent probability distributions) of $\beta_c$, $\gamma_c$, and $R_0$, for $N \in [93, 186]$. The bottom panel shows the ICC curve for the optimal parameter values, corresponding to a value of $R_0 = 1.52$.}
\end{figure}

We use the same data set as in \cite{Lega16}, estimated from the epidemiological curve provided in \cite{Fernandez08}. We fit the ICC curve to the interpolated data (i.e. find $\beta_c(N)$, $\gamma_c(N)$, and $p_c(N)$) for a range of values of $N$, identify the value $N_m$ of $N$ that best fits the data (i.e. for which the RMSE is minimum), and select a range of values of $N$ near $N_m$ that give a RMSE within 2\% of its minimum. Then, for each value of $N$ in the selected range, we apply the parameter estimation procedure of Section \ref{sec:params}, with 10,000 oubreak realizations, obtained from adding Poisson-distributed noise to the smoothed and interpolated data. Figure \ref{figure:Gastro} shows the resulting parameter distributions (top panel; $N \in [93, 186]$) and a plot of the ICC curve for the optimal parameter values (bottom panel; $N = 120$, $\beta = 1.34$, and $\gamma = 0.88$). The value of $R_0$ for the ICC curve displayed is $1.52$ and corresponds to an attack rate of 59.8\%. This is near the upper boundary of the 95\% confidence interval (38.5 \% - 61.3 \%) for the overall attack rate, but well within the 95\% confidence interval of the attack rate among nursing home residents (42.1 \% - 72.2 \%) reported in \cite{Fernandez08}.

\subsection{COVID-19}
\begin{figure}[t]
\includegraphics[width=\linewidth]{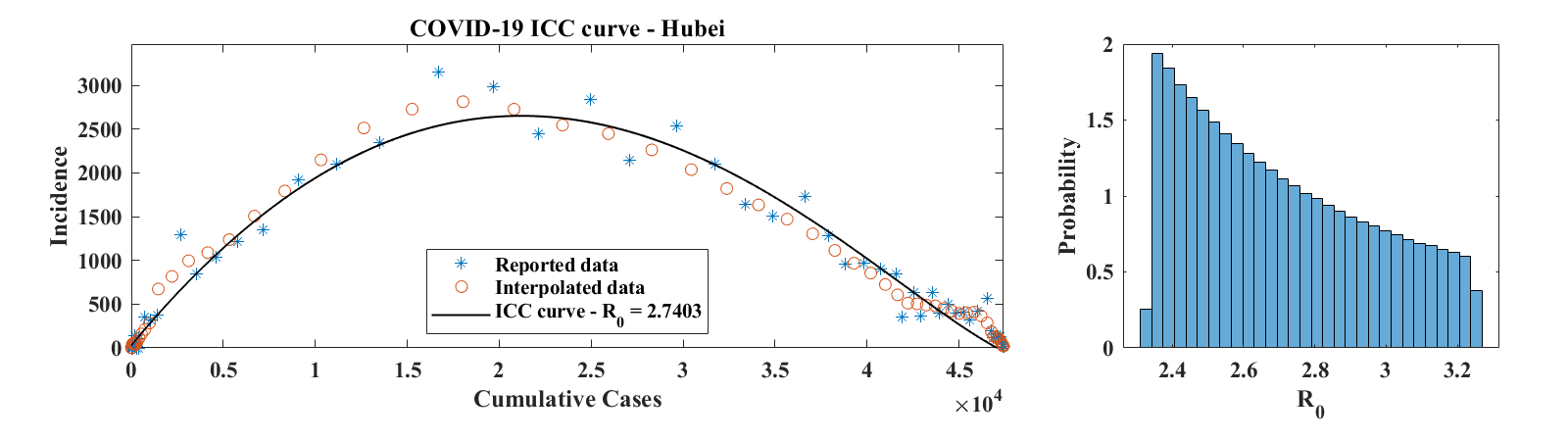} \\
\medskip
\includegraphics[width=\linewidth]{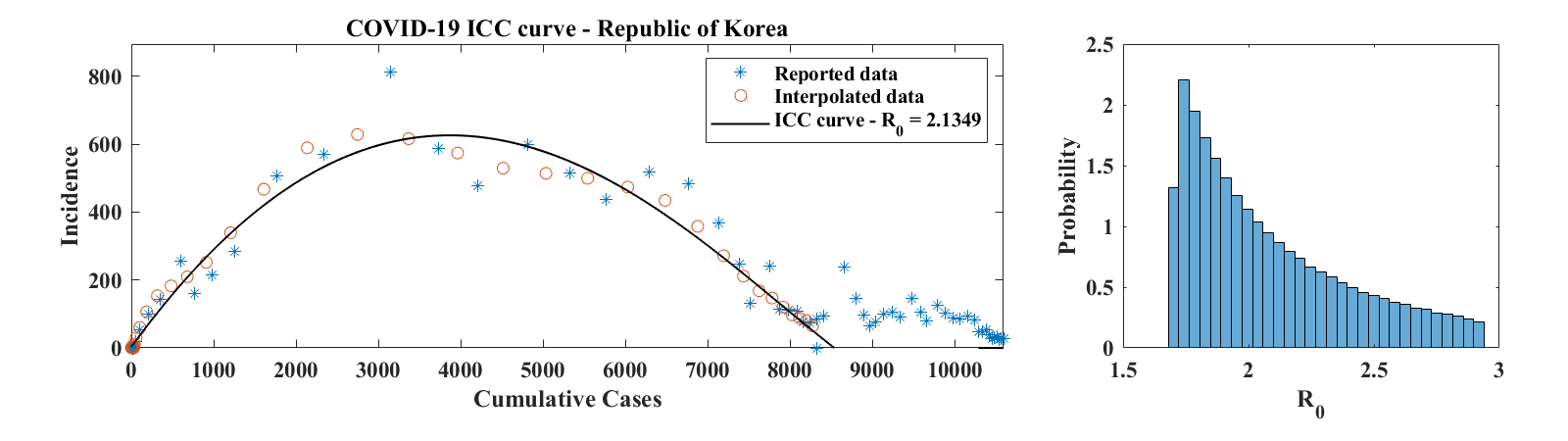} \\
\medskip
\includegraphics[width=\linewidth]{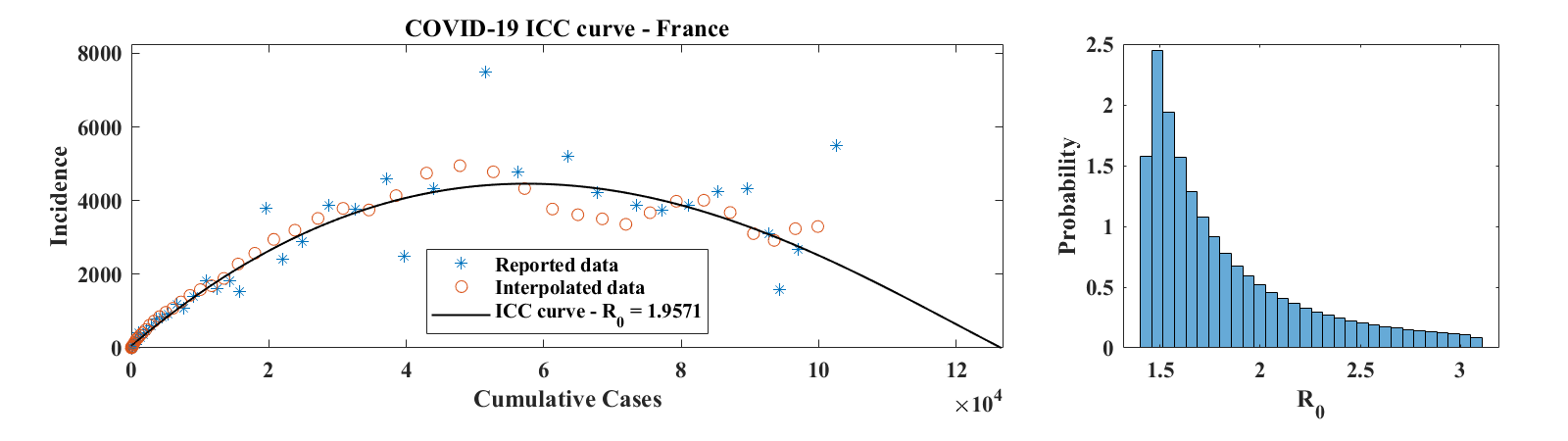}
\caption{\label{figure:COVID} (Color online). Optimal ICC curves and estimated distributions of $R_0$ values for the COVID-19 outbreaks in Hubei Province (top panel), the Republic of Korea (middle panel), and France (bottom panel).}
\end{figure}

Figure \ref{figure:COVID} shows ICC curves for COVID-19 outbreaks in a few countries. The data was obtained from the World Health Organization daily situation reports \cite{WHO} as well as from the Wuhan Health Municipal Commission \cite{WHMC}. For Hubei Province, only laboratory confirmed cases were used. Consequently, daily increment values for 2/17-19/2020 (days 65 to 67 in the data set, when China combined laboratory and clinically confirmed cases in its reports for a brief period of time) were set to half of the reported number of confirmed cases; the number of cumulative cases was then obtained by summing daily reports of laboratory confirmed cases. For the Republic of Korea, only the first 58 points in the data set were used, since they correspond to the first wave of the outbreak. The third example is for a country (France), where the outbreak was not over as of 4/15/2020. For each of these examples, we show the ICC curve for the set of parameters that best fits the interpolated data. The optimal values of $N$ are $N = 51160$ for Hubei Province, $N = 10282$ for the Republic of Korea, and $N = 161142$ for France. The corresponding optimal basic reproduction ratio values are $R_0 = 2.74$ for Hubei Province, $R_0 = 2.13$ for the Republic of Korea, and $R_0 = 1.96$ for France. A histogram of estimated values of $R_0$, scaled to represent a probability distribution function, is shown in the right panel of each row of Figure \ref{figure:COVID}. These were obtained as described above in the case of the outbreak in Mallorca, except that no Poisson-distributed noise was added to the data. This is because incidence values are generally large and that, as discussed in Section \ref{sec:params}, the relative effect of the added noise is minimal, leading to a level of variability of $R_0$ that is insignificant compared to the effect of changing $N$. Values of $R_0$ in the 2-3 range are consistent with early estimates of the basic reproduction ratio of COVID-19 published in the literature \cite{Wu20,MIDAS}.

\begin{figure}[h]
\includegraphics[width=\linewidth]{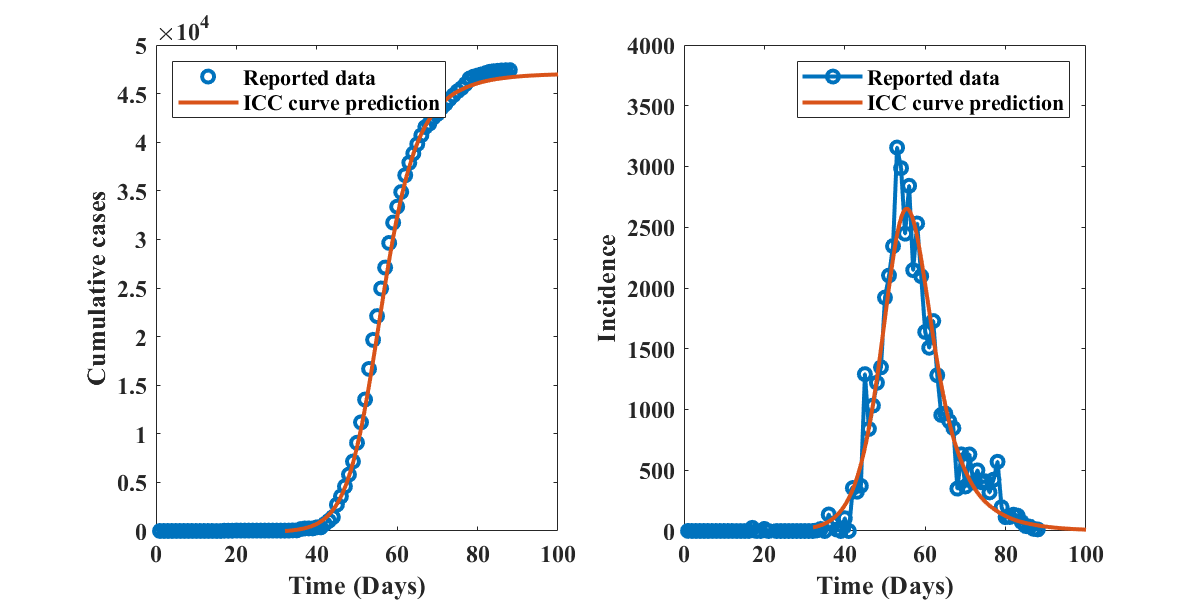} \\
\caption{\label{figure:SIR_Hubei} (Color online). Cumulative cases (left) and incidence (right) as functions of time for the COVID-19 outbreak in Hubei Province, China. The solid curves represent the predictions of the optimal ICC curve and the open circles are the data points. Time is measured in days from 12/14/2019 (day 0).}
\end{figure}

The time course of the COVID-19 outbreak in Hubei Province, as described by the SIR model with optimal parameters identified by the present method ($\beta = 0.401$, $\gamma = 0.146$ ($R_0 = 2.74$), and $N = 51160$) is shown in Figure \ref{figure:SIR_Hubei}, together with the reported data. The solid curves are obtained by integration of the ICC curve, with initial conditions corresponding to day 10 of the outbreak, and alignment with the data at day 45. Similar plots for the other outbreaks discussed in this section are provided as supplementary material. The good visual agreement between simulation and data indicates that the SIR model, and therefore the ICC curve approach presented in this manuscript, are able to capture the overal dynamics of real-life outbreaks.

\section{Conclusions}
\label{sec:conclusions}

This article introduces a parameter estimation method for disease outbreaks that bypasses the numerical integration of epidemiological models. The approach relies on the use of ICC curves, also introduced here. ICC curves relate incidence to the cumulative number of cases $C$ and may be viewed as nonlinear transformations of the traditional epidemiological curves, in which the time variable is replaced by $C$, a monotonically increasing but nonlinear function of time. For each single-wave outbreak, the ICC curve has a simple concave-down shape that crosses the horizontal axis at the origin and at the final value of $C$.

The formulas presented in Section \ref{sec:ICC_curves}, which extend the parabolic approximation suggested in \cite{Lega16}, are exact for the SIR model. The numerical experiments of Sections \ref{sec:params} and \ref{sec:convergence} show the method is robust to noise and may be used for parameter estimation as an outbreak unfolds, with the understanding that reasonably accurate estimates can only be reached shortly before, or after, incidence peaks. This is not a shortcoming of the present approach and was also observed in \cite{Tuncer18} when fitting synthetic prevalence time series. 

Because of the simplicity of Equation \ref{eq:ICC}, and the existence of a unique set of parameters that best fits any collection of data points, parameter distributions may be generated directly from epidemiological data. In the case of the SIR model, this methodology presents a fast and novel alternative to more traditional parameter estimation strategies, such as particle Makov Chain Monte Carlo methods (PMCMC; see \cite{Endo19} for a review). This may be beneficial in pandemic situations where epidemiological estimates need to be updated daily and in many locations simultaneously. For instance, recent work on COVID-19 data from Wuhan suggests that an SIR model can better capture the information contained in case reports than an SEIR model \cite{Roda20}. For more complex compartmental models, distributions generated by the present approach may be used as priors for PMCMC estimations of the contact and recovery rates of a disease.

When applying the proposed method to surveillance data, an estimate of $N$ may not always be available. The examples of Section \ref{sec:examples} suggest that this parameter may be identified by optimizing the fit between the ICC curve and a smoothed and interpolated version of the reported data. However, any variability in the selected value of $N$ will be associated with variability in estimates of the basic reproduction number $R_0$. Indeed, for an outbreak that has completed its course, any good fit of the reported data by an ICC curve will produce consistent values of $C_\infty$, the final number of cases for the outbreak (see for instance the plots of Figure \ref{figure:COVID}). Because of Equation \ref{eq:R0}, if $C_\infty$ is known, the estimated value of $R_0$ is only a function of $N$. This uncertainty is inherent to the SIR model and cannot be avoided. The existence of a value $N_m$ of $N$ that best fits the data is encouraging, but more robust estimates are likely to be obtained if additional information is available. In the absence thereof, a range of values of $N$ close to the optimal value $N_m$ should be used to produce credible intervals for the basic reproduction ratio $R_0$.

As previously mentioned, empirical (see \cite{Lega16}) and numerical observations by the author suggest the concave-down shape of the ICC curve is ubiquitous in outbreak data and in compartmental epidemiological models. It would therefore be interesting to explore methods which, like the discussion of Section \ref{sec:ICC_curves}, lead to exact formulations of ICC curves. In particular, knowledge of how model parameters affect the shape of ICC curves could provide simple means to visualize the consequences of mitigation effects. Combined with data assimilation, ICC curves may thus present a convenient paradigm for forecasting the course of an outbreak.

\vfill \eject
\noindent{\bf Declaration of competing interests.} \\
No potential competing interest was reported by the author.

\bigskip
\noindent{\bf Data availability statement.} \\
For codes and data sets used in this study, please see \url{https://jocelinelega.github.io/EpiGro/}.

\bigskip
\appendix
\section{Critical parameter values}
\label{AppA}
The function ${\mathcal E}_e$ defined in Equation \eqref{eq:cost} has a unique extremizer $(\beta_c, \gamma_c, p_c)$ given by the following expressions:
\begin{align*}
\beta_c &= \frac{(- B_s O_s + F_s A_s) p_c^2 + (2 O_s D_s - F_s N_s -  A_s Q_s) p_c + (N_s Q_s - O_s P_s)}{(A_s^2 - B_s L_s) p_c^2 + 2 (- A_s N_s + D_s L_s) p_c - L_s P_s + N_s^2},\\
\gamma_c &= \frac{(- A_s O_s + F_s L_s) p_c - L_s Q_s + N_s O_s}{(A_s^2 - B_s L_s) p_c^2 + 2 (- A_s N_s + D_s L_s) p_c - L_s P_s + N_s^2},\\
p_c &= \frac{(-N_s Q_s + O_s P_s) A_s + (L_s Q_s - N_s O_s) D_s + (-L_s P_s + N_s^2) F_s}{(A_s O_s - F_s L_s) D_s - (A_s^2 - B_s L_s) Q_s - N_s (B_s O_s - F_s A_s)},
\end{align*}
where
\begin{align*}
A_s & =\sum_{i=1}^M \frac{C_i}{N} \left(1 - \frac{C_i}{N}\right)^2 = \sum_{i=1}^M U_i \, P_i, \qquad
B_s = \sum_{i=1}^M \left(1-\frac{C_i}{N}\right)^2 = \sum_{i=1}^M P_i^2,\\
D_s & = \sum_{i=1}^M \ln\left(1 - \frac{C_i}{N}\right) \left(1 - \frac{C_i}{N}\right)^2 = - \sum_{i=1}^M V_i P_i , \qquad
F_s =\sum_{i=1}^M \frac{I_i}{N} \left(1 - \frac{C_i}{N}\right) = \sum_{i=1}^M {\mathcal J}_i \, P_i,\\
L_s &= \sum_{i=1}^M \left(\frac{C_i}{N}\right)^2 \left(1 - \frac{C_i}{N}\right)^2 = \sum_{i=1}^M U_i^2 , \qquad 
N_s = \sum_{i=1}^M  \left(1 - \frac{C_i}{N}\right)^2 \ln\left(1 - \frac{C_i}{N}\right) \frac{C_i}{N} = - \sum_{i=1}^M U_i\, V_i,\\
O_s & = \sum_{i=1}^M \frac{I_i}{N} \left(1 - \frac{C_i}{N}\right) \frac{C_i}{N} = \sum_{i=1}^M {\mathcal J}_i\, U_i, \qquad
P_s = \displaystyle \sum_{i=1}^M \left(\ln\left(1 - \frac{C_i}{N}\right)\right)^2 \left(1- \frac{C_i}{N}\right)^2 = \sum_{i=1}^M V_i^2,\\
Q_s & =\sum_{i=1}^M \frac{I_i}{N} \ln\left(1 - \frac{C_i}{N}\right) \left(1 - \frac{C_i}{N}\right) =- \sum_{i=1}^M {\mathcal J}_i \, V_i,
\end{align*}
and we have used the notation
\[
{\mathcal J}_i = \frac{I_i}{N}, \quad P_i = 1-\frac{C_i}{N}, \quad U_i = \frac{C_i}{N} \left(1- \frac{C_i}{N}\right), \quad V_i = - \ln \left(1-\frac{C_i}{N}\right) \, \left(1-\frac{C_i}{N}\right).
\]

\vfill \eject

\end{document}


\maketitle

We show below how the distributions of $\beta_c$, $\gamma_c$, and the basic reproductive number $R_0 = \frac{\beta_c}{\gamma_c}$ defined in the main body of the article evolve as functions of time during the course of an outbreak. We consider two examples, with values of $R_0$ equal to 2 and 10.5 respectively. For the former, Figure \ref{figsup:convergence_P_2} shows how histograms (scaled to approximate distribution functions) of parameter estimates in the presence of Poisson-distributed noise narrow and center on the actual parameter values as the number of available data points increases. Figure \ref{figsup:convergence_G_2} shows similar results for normally-distributed noise of amplitude 0.15. The corresponding time-dependence of $S$, $I$, and $R$ is displayed in the left panel of Figure 1, and the temporal evolution of the means and standard deviations of the parameter estimates are shown in Figure 6, both in the main article.

\begin{figure}[h]
\vskip 1.truecm
\includegraphics[width=\linewidth]{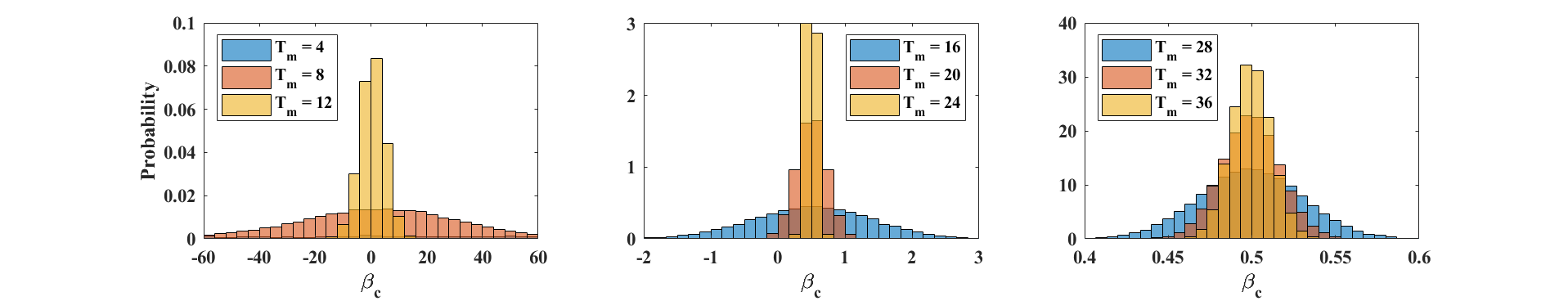} \\
\includegraphics[width=\linewidth]{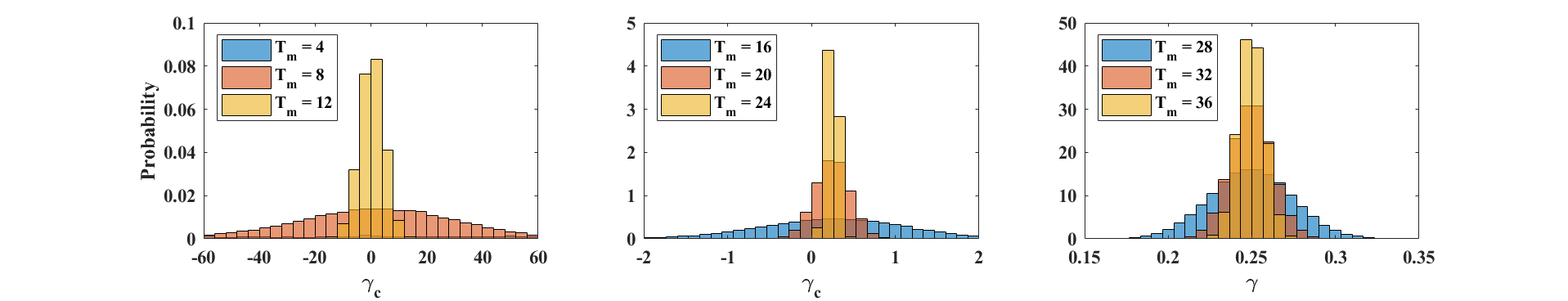} \\
\includegraphics[width=\linewidth]{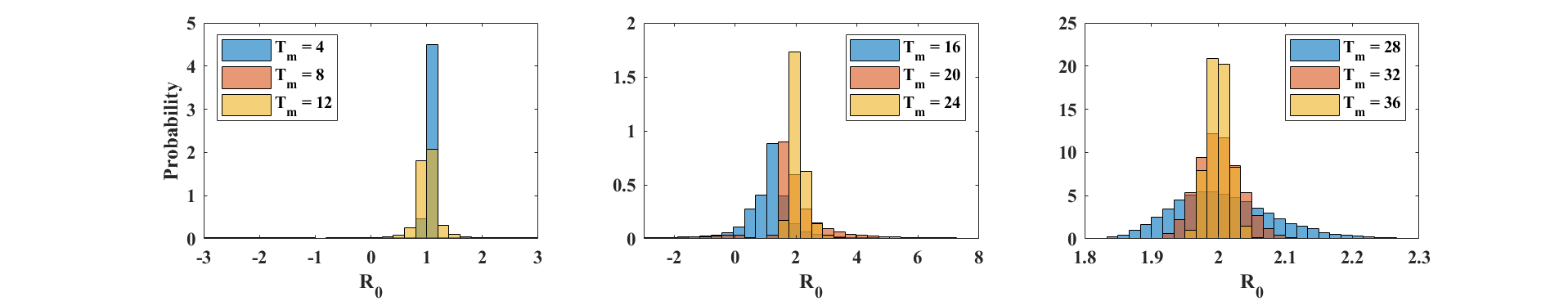}
\caption{\label{figsup:convergence_P_2} Temporal evolution of the distributions of $\beta_c$, $\gamma_c$ and $R_0$ as the outbreak unfolds, for Poisson-distributed noise in the case of an outbreak with $R_0 = 2$.}
\end{figure}

\begin{figure}[t]
\includegraphics[width=\linewidth]{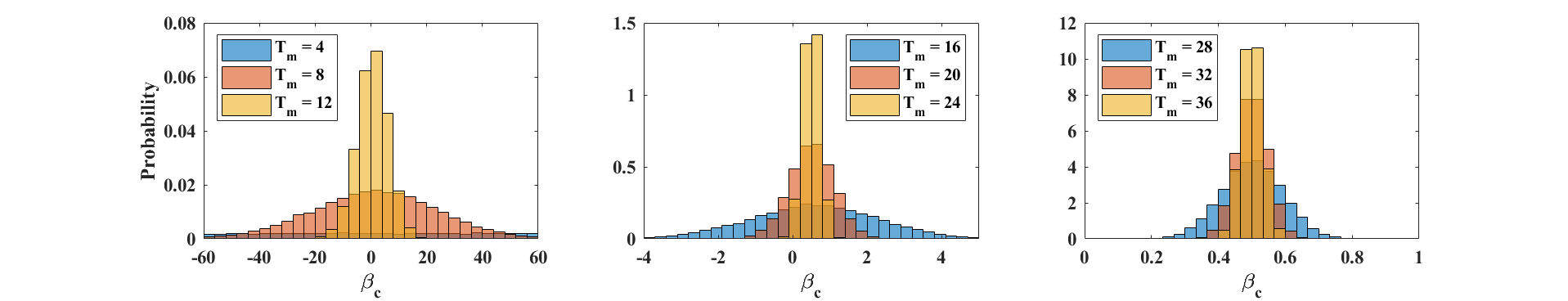} \\
\includegraphics[width=\linewidth]{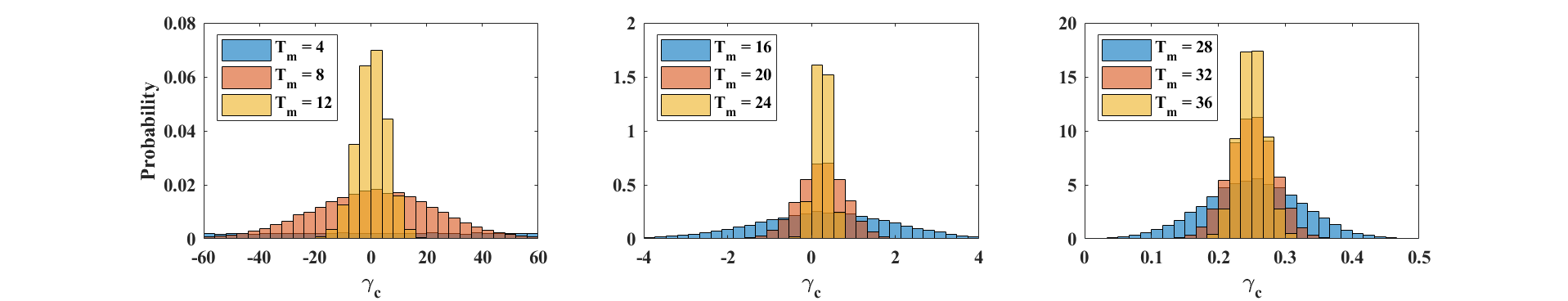} \\
\includegraphics[width=\linewidth]{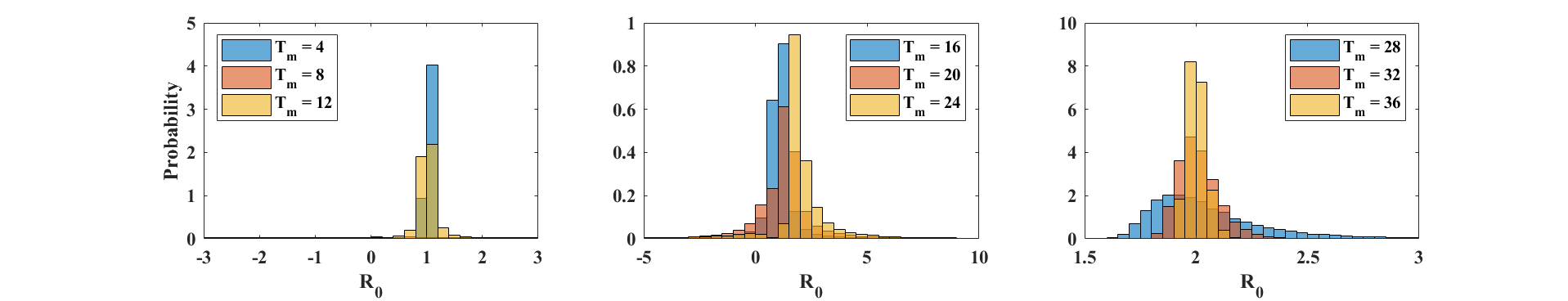}
\caption{\label{figsup:convergence_G_2} Temporal evolution of the distributions of $\beta_c$, $\gamma_c$ and $R_0$ as the outbreak unfolds, for normally-distributed noise of amplitude 0.15, in the case of an outbreak with $R_0 = 2$.}
\end{figure}

\vfill \eject
\newpage
For $R_0 = 10.5$, the time course of $S$, $I$, and $R$ (in the absence of noise) together with the corresponding ICC curve, are shown in Figure \ref{figsup:SIR_simulations}. The following pages show the temporal evolution of the histograms of $\beta_c$, $\gamma_c$, and $\beta_c / \gamma_c$ for Poisson-distributed noise (Figure \ref{figsup:convergence_P_10_5}) and normally-distributed noise of amplitude 0.05 (Figure \ref{figsup:convergence_G_10_5_05}) and 0.15 (Figure \ref{figsup:convergence_G_10_5_15}). Finally, Figure \ref{figsup:convergence_a} and Figure \ref{figsup:convergence_b} show the temporal evolution of the parameter distribution means, together with error bars of one standard deviation about the mean.

\begin{figure}[t]
\vskip .7 truecm
\includegraphics[width=\linewidth]{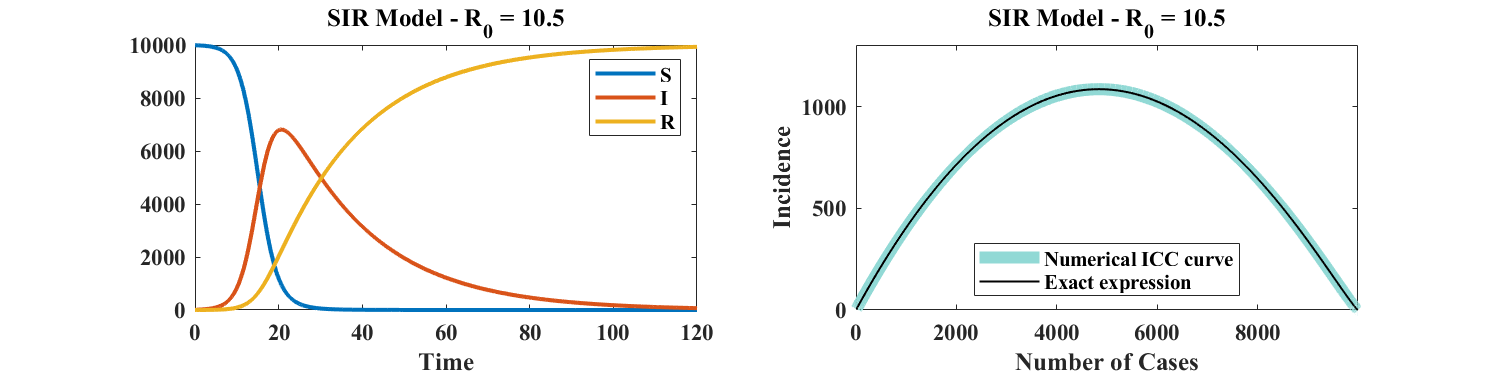}
\caption{\label{figsup:SIR_simulations} Simulation of the SIR model showing the time-dependence of the S, I, and R compartments (left), for $R_0 = 10.5$. The corresponding ICC curve is shown on the right. The exact expression given by Equation (1) of the main article is plotted as a solid black line; the numerically estimated ICC curve is shown as a  thick cyan line.}
\end{figure}

\begin{figure}[t]
\includegraphics[width=\linewidth]{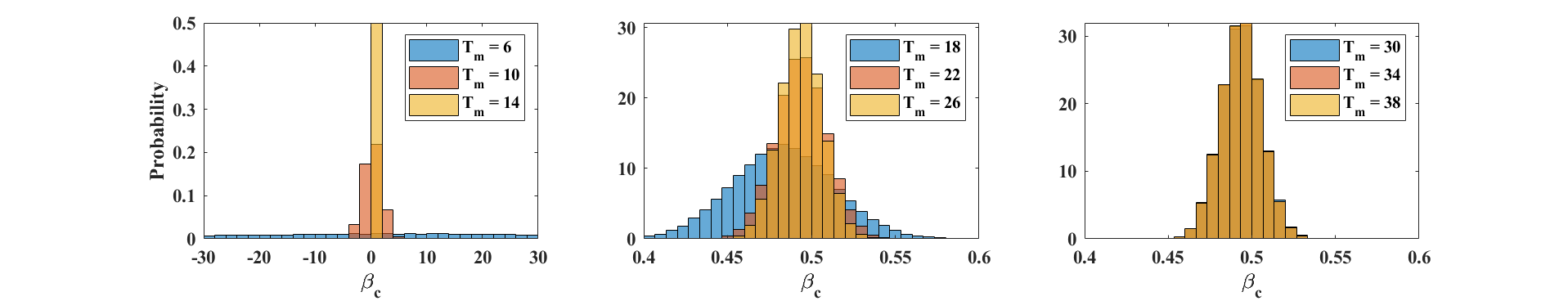} \\
\includegraphics[width=\linewidth]{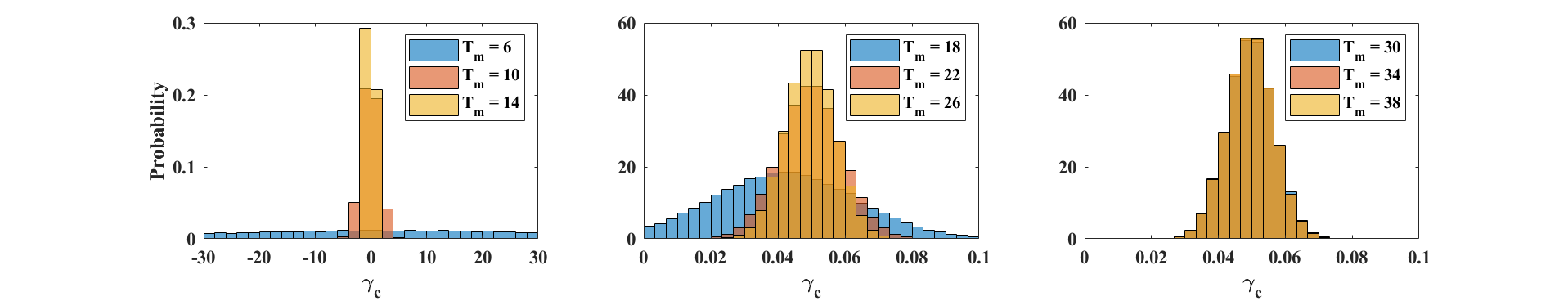} \\
\includegraphics[width=\linewidth]{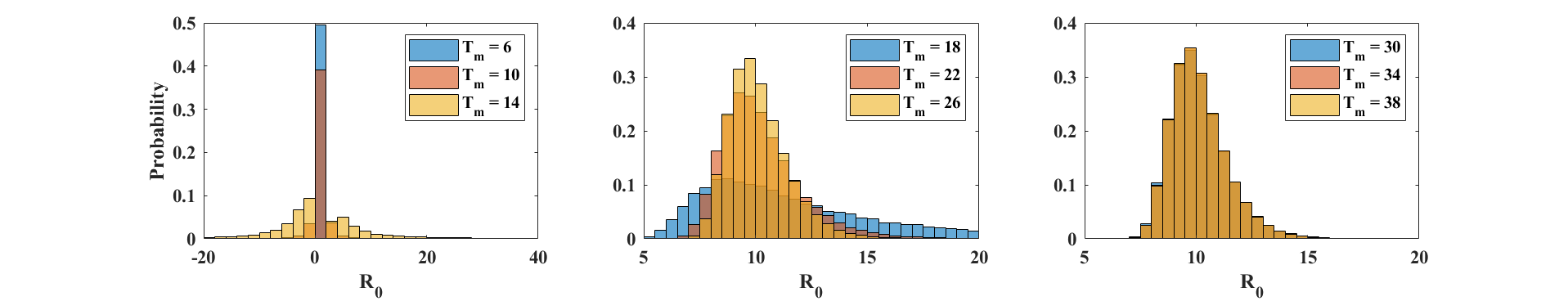}
\caption{\label{figsup:convergence_P_10_5} Temporal evolution of the distributions of $\beta_c$, $\gamma_c$ and $R_0$ as the outbreak unfolds, for Poisson-distributed noise in the case of an outbreak with $R_0 = 10.5$.}
\end{figure}

\begin{figure}[t]
\includegraphics[width=\linewidth]{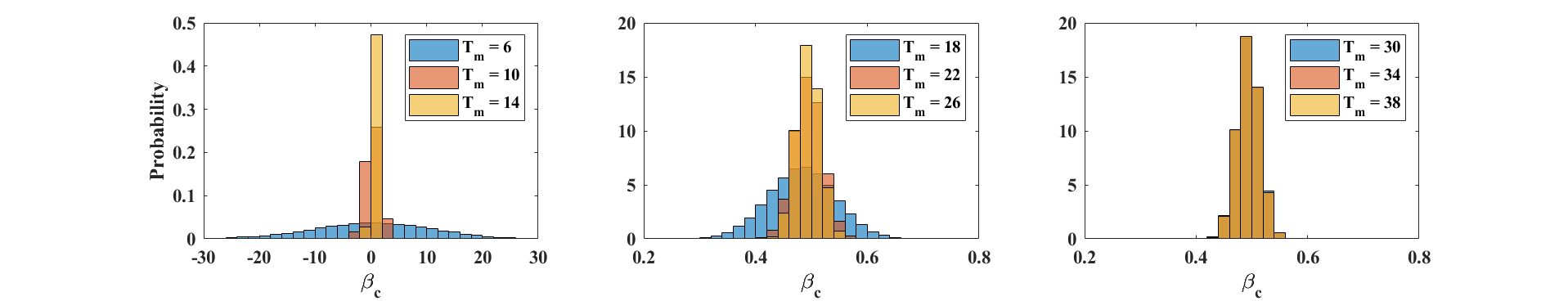} \\
\includegraphics[width=\linewidth]{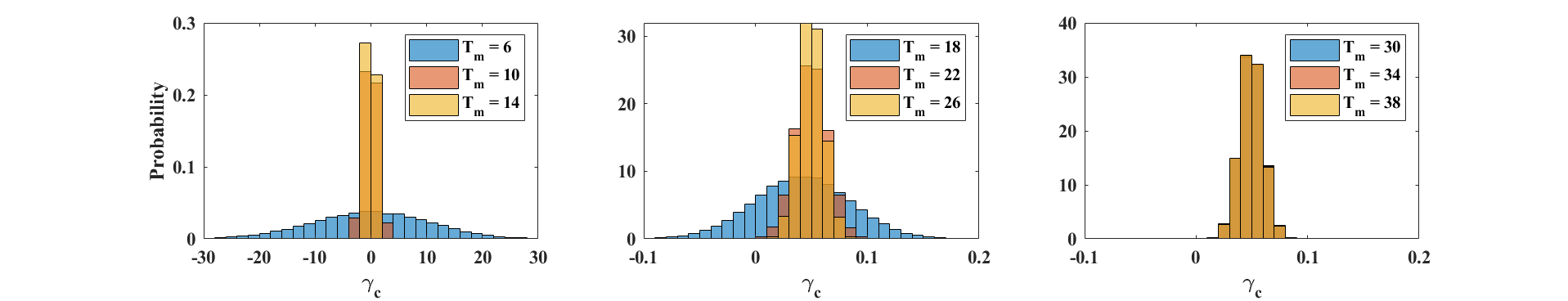} \\
\includegraphics[width=\linewidth]{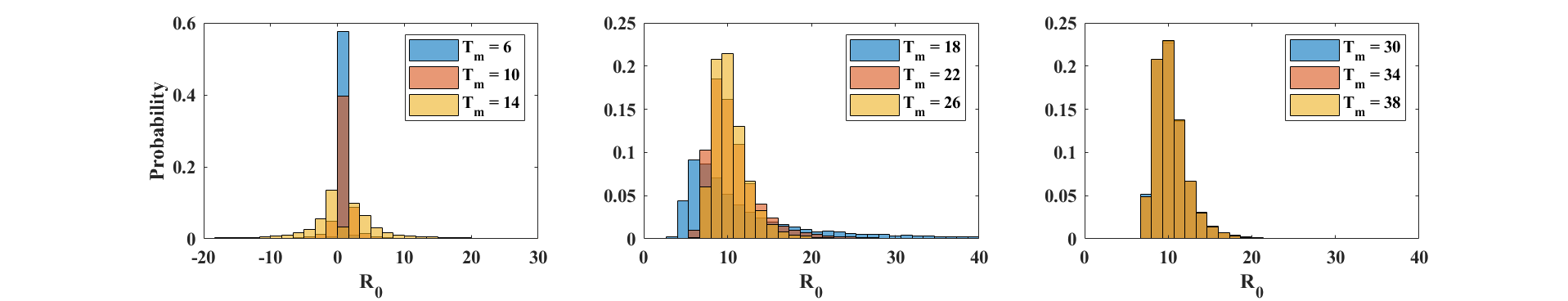}
\caption{\label{figsup:convergence_G_10_5_05} Temporal evolution of the distributions of $\beta_c$, $\gamma_c$ and $R_0$ as the outbreak unfolds, for normally-distributed noise of amplitude 0.05, in the case of an outbreak with $R_0 = 10.5$.}
\end{figure}

\begin{figure}[t]
\includegraphics[width=\linewidth]{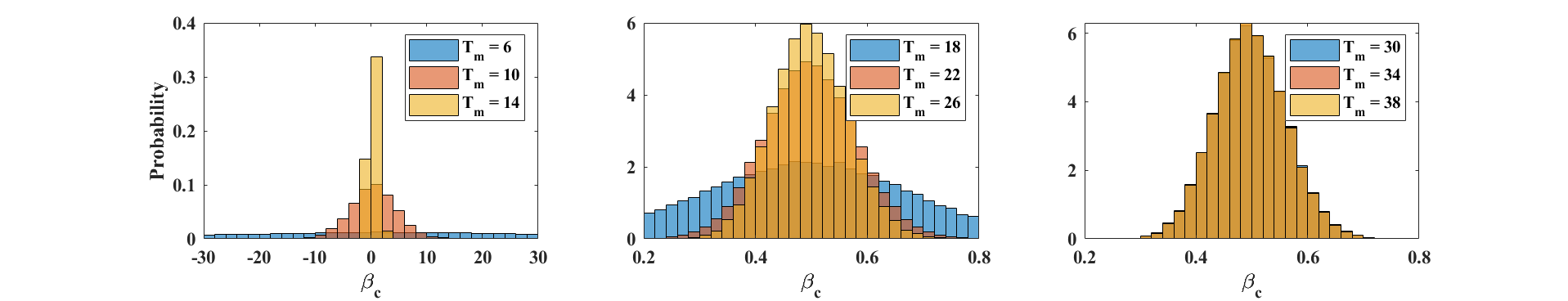} \\
\includegraphics[width=\linewidth]{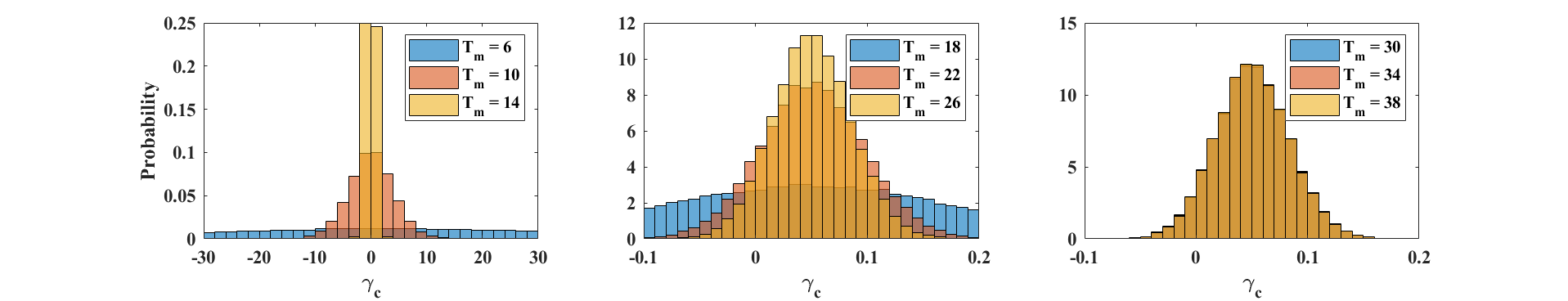} \\
\includegraphics[width=\linewidth]{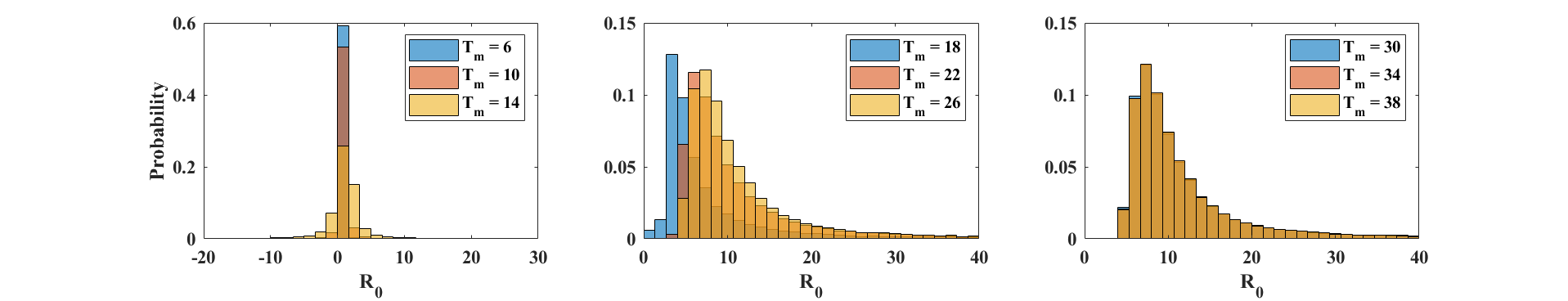}
\caption{\label{figsup:convergence_G_10_5_15} Temporal evolution of the distributions of $\beta_c$, $\gamma_c$ and $R_0$ as the outbreak unfolds, for normally-distributed noise of amplitude 0.15, in the case of an outbreak with $R_0 = 10.5$.}
\end{figure}

\begin{figure}[t]
\vskip 1.truecm
\includegraphics[width=0.48 \linewidth]{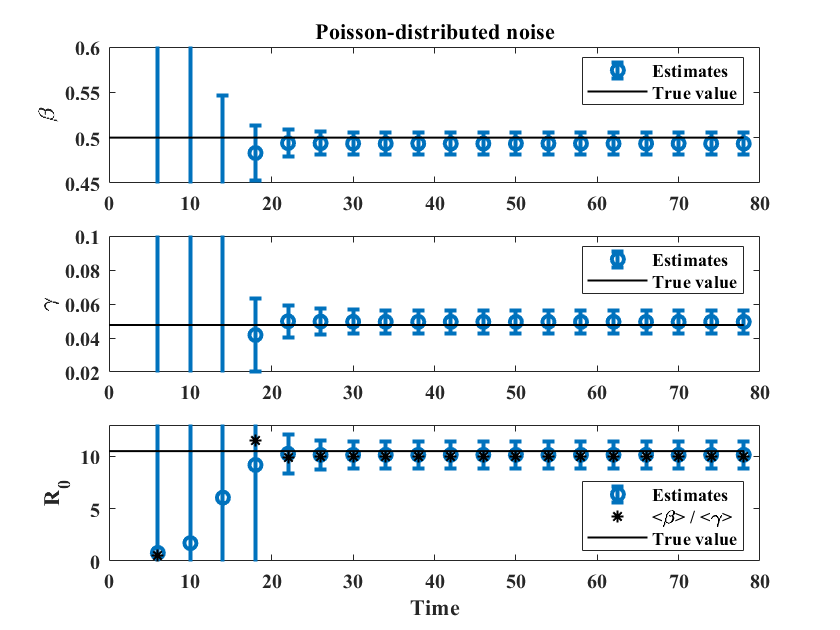}
\includegraphics[width=0.48 \linewidth]{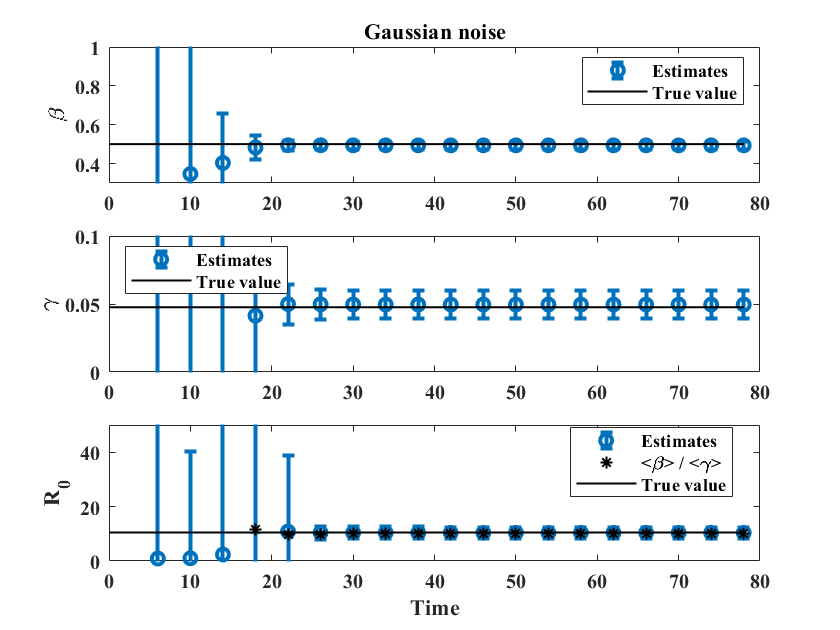}
\caption{\label{figsup:convergence_a} Estimates of $\beta$, $\gamma$ and $R_0$ as the outbreak unfolds for Poisson-distributed noise (left) and normally-distributed noise of size 0.05 (right), and with $R_0 = 10.5$. Estimates of $R_0$ show the evolution of $<\beta / \gamma>$ (circles) and $<\beta> / <\gamma>$ (stars), where $< \cdot >$ indicates sample mean. Error bars correspond to one standard deviation on each side of the mean.}
\end{figure}

\begin{figure}[t]
\includegraphics[width=0.48 \linewidth]{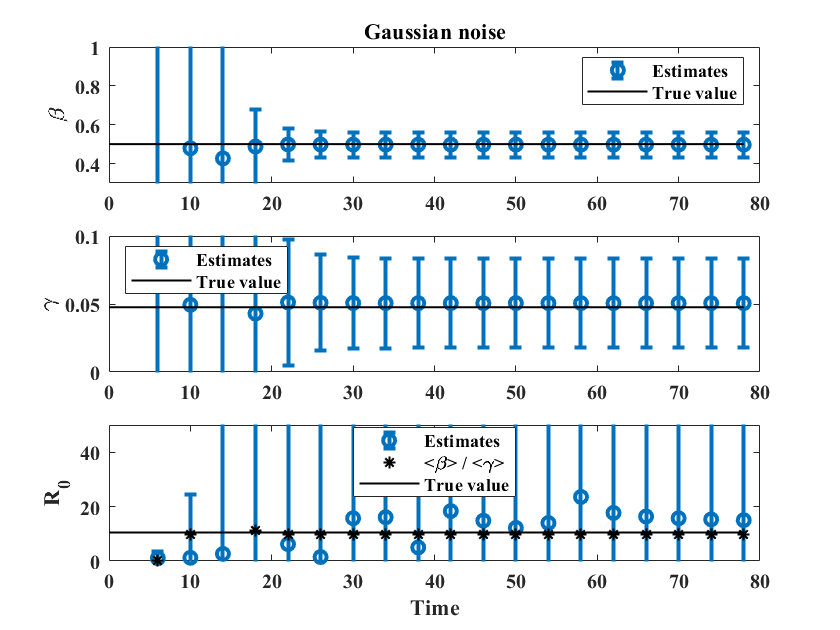}
\caption{\label{figsup:convergence_b} Estimates of $\beta$, $\gamma$ and $R_0$ as the outbreak unfolds for normally-distributed noise of size 0.15, and with $R_0 = 10.5$. Estimates of $R_0$ show the evolution of $<\beta / \gamma>$ (circles) and $<\beta> / <\gamma>$ (stars), where $< \cdot >$ indicates sample mean. Error bars correspond to one standard deviation on each side of the mean.}
\end{figure}


\maketitle

\begin{figure}[h]
\includegraphics[width=\linewidth]{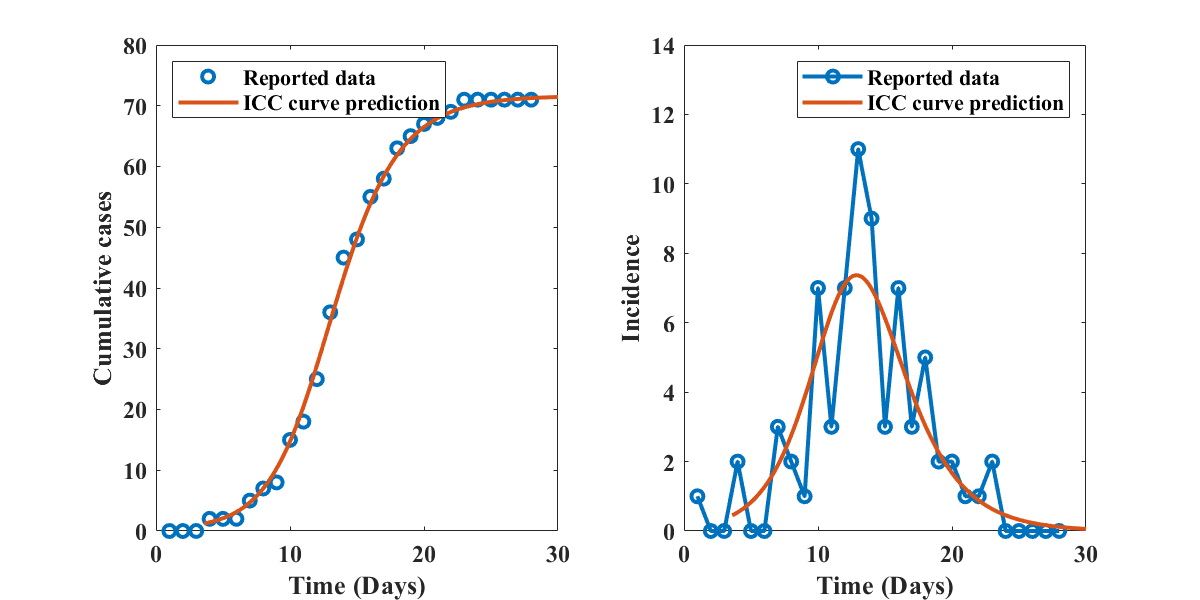} \\
\caption{\label{figsup:SIR_Gastro} Cumulative cases (left) and incidence (right) as functions of time for the gastroenteritis outbreak in a nursing home in Mallorca, Spain. The solid curves represent the predictions of the optimal ICC curve with averaged parameter values $\beta = 1.338$, $\gamma=0.879$ ($R_0 = 1.523$), and $N = 120$, and the open circles are the data points.}
\end{figure}

\begin{figure}[t]
\includegraphics[width=\linewidth]{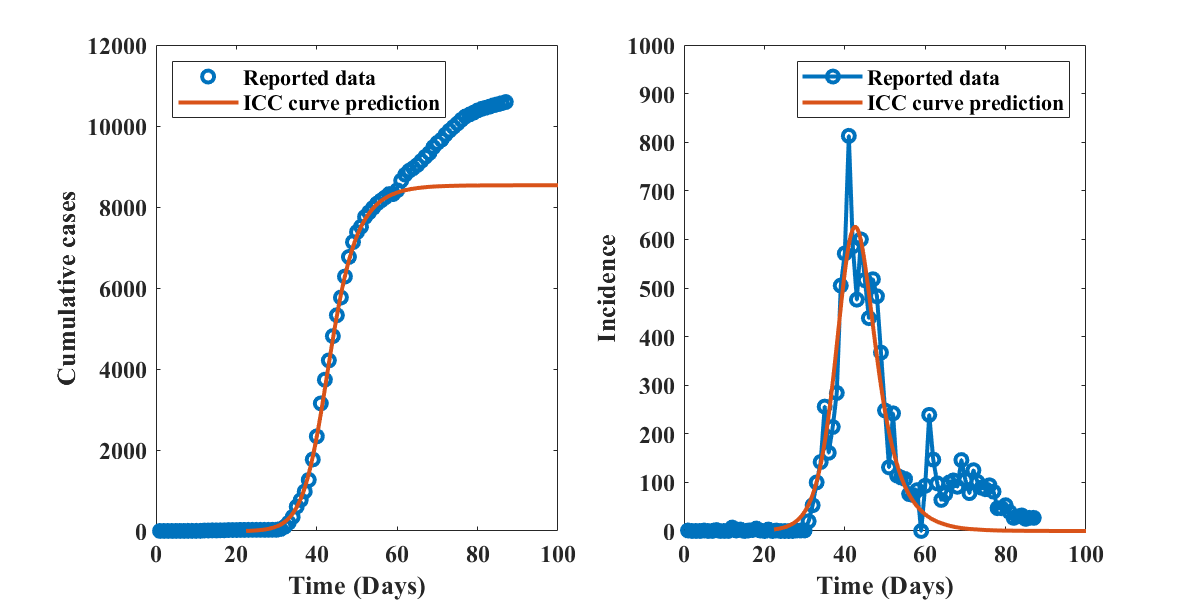} \\
\caption{\label{figsup:SIR_Korea} Cumulative cases (left) and incidence (right) as functions of time for the COVID-19 outbreak in the Republic of Korea. The solid curves represent the predictions of the optimal ICC curve with parameters $\beta = 0.627$, $\gamma=0.294$ ($R_0 = 2.135$), and $N = 10282$, and the open circles are the data points. Time is measured in days from 1/19/2020 (Day 0). The ICC curve approximation does not consider the second wave of the outbreak, starting around Day 60.}
\end{figure}

\begin{figure}
\includegraphics[width=\linewidth]{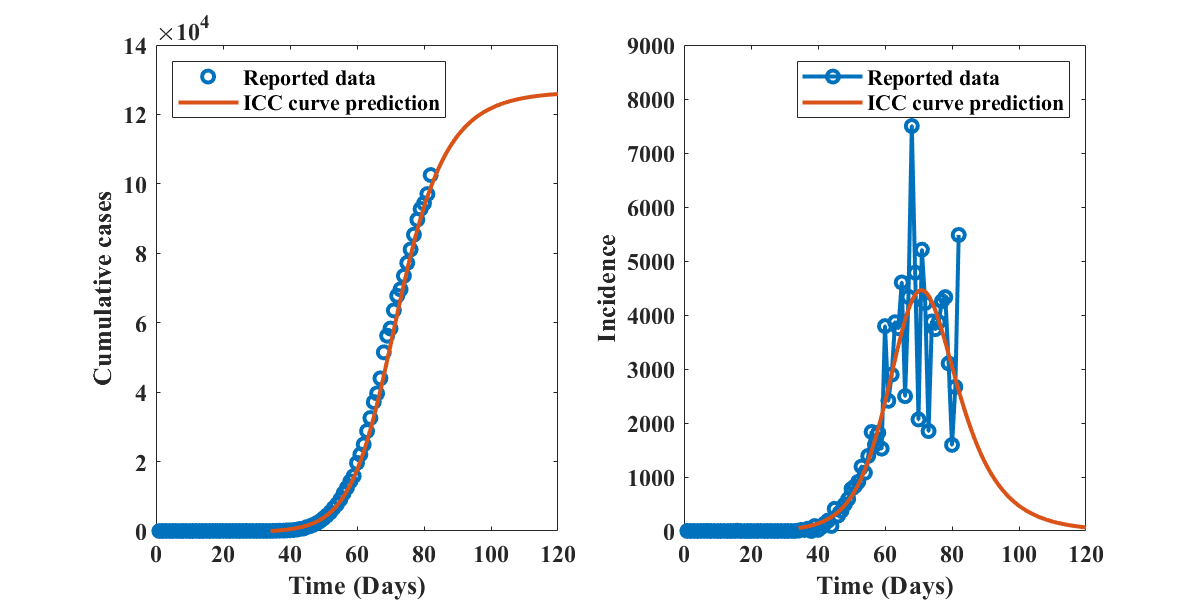} \\
\caption{\label{figsup:SIR_France} Cumulative cases (left) and incidence (right) as functions of time for the COVID-19 outbreak in France. The solid curves represent the predictions of the optimal ICC curve with parameters $\beta = 0.325$, $\gamma=0.166$ ($R_0 = 1.957$), and $N = 161138$, and the open circles are the data points. Time is measured in days from 1/24/2020 (Day 0).}
\end{figure}

Figures \ref{figsup:SIR_Gastro}, \ref{figsup:SIR_Korea}, and \ref{figsup:SIR_France} compare the time course of the gastroenteritis outbreak in Mallorca, Spain \footnote{M.A. Luque Fern{\'a}ndez, A. Galm{\'e}s Truyols, D. Herrera Guibert, G. Arbona Cerd{\'a}, F. Sancho Gay{\'a}, {\it Cohort study of an outbreak of viral gastroenteritis in a nursing home for elderly, Majorca, Spain, February 2008}, Euro Surveill. {\bf 13(51)}, pii=19070 (2008). \href{https://www.eurosurveillance.org/content/10.2807/ese.13.51.19070-en}{https://www.eurosurveillance.org/content/10.2807/ese.13.51.19070-en}} and of the COVID-19 outbreaks in the Republic of Korea and in France, to the predictions of the ICC curves discussed in the main article. 

